# A hierarchical life cycle model for Atlantic salmon stock assessment at the North Atlantic basin scale

Etienne RIVOT [1,2,*], Maxime OLMOS [1,2,#]

Gérald CHAPUT [3], Etienne PREVOST [4,2]

[1] ESE, Ecology and Ecosystems Health, Agrocampus Ouest, INRA, Rennes, France

[2] Management of Diadromous Fish in their Environment, AFB, INRA, Agrocampus Ouest, E2S UPPA, Rennes, France

[3] Fisheries and Oceans, Moncton, Canada

[4] INRA, UMR 1224 Ecobiop, INRA, UPPA, Aquapôle, St Pée sur Nivelle, France

*Email: etienne.rivot@agrocampus-ouest.fr

#Email: maxime.olmos@agrocampus-ouest.fr

# Foreword and acknowledgements

**This working paper is derived from:**

**Aknowledgements**

This study was made possible by the work of the numerous people who collect and compile the data used by the ICES Working Group on North Atlantic Salmon. For North America: Dave Reddin, Martha Robertson and Geoff Veinott of Fisheries and Oceans Canada (DFO) Newfoundland and Labrador Region, Ross Jones and Alex Levy of DFO Maritimes Region, Cindy Breau, Michel Biron, Scott Douglas of DFO Gulf Region, Mélanie Dionne and Julien April from the provincial government of Quebec, and Tim Sheehan and Joan Trial of USA. For Southern NEAC: Dennis Ensing (UK Northern Ireland), Gudni Gudbergsson (Iceland), Michael Millane and Niall O Maoiléidigh (Ireland), Ted Potter and Ian Russel (UK England and Wales), Gordon Smith (UK Scotland), and the Pôle AFB-IRA Gest-Aqua lead by Laurent Beaulaton (France). We thank Jerome Guitton for his help in the map design and Maud Queroue for her help gathering the freshwater data.

The project has received funding from the European Union's Seventh Framework Programme (FP7/2007– 2013) under grant agreement No. 244706/ECOKNOWS project, from the Agence Française de la Biodiversité under grant agreement INRA-AFB SalmoGlob 2016-2018, and by the European Regional Development Fund through the Interreg Channel VA Programme, project SAMARCH Salmonid Management Round the Channel.

# Table of contents

# Abstract

We developed an integrated hierarchical Bayesian life cycle model that simultaneously estimates the abundance of post-smolts at sea, post-smolt survival rates, and proportions maturing as 1SW, for all SU in Northern Europe, Southern Europe and North America. The model is an age- and stage-based life cycle model that considers 1SW and 2SW life history strategies and harmonizes the life history dynamics among SU in North America and Europe. The new framework brought a major contribution to improve the scientific basis for Atlantic salmon stock assessment. It is a benchmark for the assessment and forecast models currently used by ICES for Atlantic salmon stock assessment in the North Atlantic.

- The model is built in a hierarchical Bayesian state-space framework that integrates both process and observation errors. Observation errors on returns and catches are integrated through a sequential approach, similar to the one developed in stock assessment models for Atlantic salmon in the Baltic. Probability distributions on returns and catches at sea were derived separately from the life cycle model (using the Run Reconstruction Models developed by ICES WGNAS, ICES 2015b) and then used to approximate likelihoods. The choice of using a sequential or a full integrated approach represents a trade-off between model realism and computational efficiency. A fully integrated model would provide a more transparent view of how the data are incorporated in the entire process being modelled. It would also allow for the option of incorporating covariation in observation errors of 1SW and 2SW annual returns. However, the currently used observation models are highly heterogeneous among SU (ICES 2015b) and developing a fully integrated model capturing all these specificities would come at a high cost of increased model complexity, and increased computational burden for Bayesian statistical inference.
- The new model constitutes an important tool for future improvement of our understanding of the mechanisms driving the response of Atlantic salmon populations to variations in biological and environmental factors in a hierarchy of spatial scales. Formulating the dynamics of all SU in a single hierarchical model provides a tool for modelling covariations among different populations that may share part of their migration routes at sea and may be exploited by the same marine fisheries. It provides a framework for quantifying the spatial coherence in the temporal variations of post-smolt survival and of the sea-age composition of returns for SU distributed across a broad gradient of longitude and latitude in the North Atlantic basin.

- Time series of marine survivals and proportions of fish maturing as 1SW are modelled as latent random walks (RW) with covariation among SUs. The RW is a simple and flexible structure for modelling trends and shifts over time. No a priori hypothesis on the sign of covariation among SU is made in our approach, and inferences on covariation and correlation among SU are derived directly from the posterior distribution of the variance-covariance matrix.
- The new life cycle model provides a singular harmonized framework to simultaneously assess two sea-classes of Atlantic salmon for all SU in North America and Europe. This represent a paradigm shift from the stock assessment and forecasting approach currently used by ICES considers the North American and European (Southern and Northern) continental stock groups separately and these models have different demographic structures (Chaput, 2012; ICES, 2015a).
- The life cycle model is a natural framework for forecasting population dynamics. The same model is used for both the inferences (hindcasting) and forecasting phases, and all the model properties are readily integrated into the forecast process: (1) All sources of uncertainty in the model (temporal variability) and the parameters (joint posterior distribution) are readily integrated in the inference and forecasting phases; (2) Temporal variations in post-smolt survivals and in the proportions of fish maturing as 1SW incorporate the covariation among SU in both the inference and forecasting phases; (3) A single model can be used to forecast the population dynamics of all SU simultaneously, which is of particular interest when assessing catch options for mixed stock fisheries operating on a mixture of stocks from both North America and Europe simultaneously. Specifically we demonstrate the use of the life cycle model to evaluate the probability that returns of spawners in all SU fall below management objectives for different catch options in both the Western Greenland and Faroes mixed stock fisheries. But the model can also be used to provide catch options for other fisheries, or to assess conservation measures for the different sea-age classes or the SU separately.

- Last, the integrated life cycle framework is expandable and provides an opportunity to assimilate various sources of information. Specifically, the life cycle model incorporates a likelihood function to assimilate genetic data to allocate catches at West Greenland among the SU, which is more realistic than the hypothesis of a homogeneous harvest rate among SU that is done in the current model used by ICES (ICES, 2015a).

The version of the model presented in this working paper was run with the data of the ICES WGNAS report 2015 (ICES, 2015a). The time series of data are therefore 44 years from 1971 to 2014.

# 1 Background

Atlantic salmon (*Salmo salar)* (hereafter A. salmon) that reproduce in rivers of eastern North America and Northeast Atlantic countries of Europe undertake wide-ranging migrations to common feeding grounds in the North Atlantic, where they are exposed to common marine environmental conditions and fisheries (Beaugrand and Reid, 2003; Beaugrand and Reid, 2012; Friedland et al., 2014; Mills et al., 2013). A. salmon are susceptible to be harvested at several stages in their life cycle. Some fisheries operate in high seas when population originating from various continental habitat regroup on high seas foraging areas, in coastal areas when salmon navigate before entering their natal river, or in freshwater (estuarine or river areas) during the final stages of their spawning migration. In particular, when present in the feeding grounds of West Greenland or in the vicinity of the Faroe Islands, they may be harvested in mixed stock fisheries, referred to as the high seas (or distant water) fisheries (Chaput, 2012; ICES, 2017a). A. salmon populations had been strongly overfished during the 1960s through the 1990s with total catch in the North Atlantic maxima of about 12 000 t in 1967 and 1973. Thus catches have participated to a decline of numbers of salmon returning to home rivers (Mills, 1989; Parrish et al., 1998). Catches at the West Greenland fishery reached a peak of just under 2700t in 1971 following the high development of offshore driftnet fishery in the 1960s (Dunbar and Thomson, 1979; Horsted, 1988).

The regulation of mixed stock high seas fishery was of sufficient concern that an international body (the North Atlantic Salmon Conservation Organization (NASCO; http://www.nasco.int/)) was formed in 1982 and a treaty subsequently signed by participating countries to manage the marine fisheries impacting different stock of A. salmon (Windsor and Hutchinson, 1994). The annual stock status reports developed by the Working Group North Atlantic Salmon of the International Council for the Exploration of the sea (ICES/CIEM WGNAS) and the subsequent scientific advices provided to the NASCO have formed the basis for the negotiations and subsequent management of these fisheries.

To manage West Greenland and Faroes fisheries, ICES provides catch advice based on a forecast of A. salmon abundance prior to the high seas fisheries exploitation (Pre

Fishery Abundance, measured at the January 1 of the first winter spent at sea, hereafter denoted PFA). A fixed escapement strategy has been adopted with the objective of achieving the spawner (or egg) requirements for the contributing stocks on both sides of the Atlantic Ocean (Chaput, 2012; Crozier et al., 2003; Crozier et al., 2004; Potter et al., 2004a).

Stock assessment models for Atlantic salmon have been developed based on data aggregated at the scale of regional or national stock units (SU) over the North Atlantic area within three continental stock groups (CSG): eastern North America (NA), Southern European (SE) and Northern European (NE) (Crozier et al., 2004; Potter et al., 2004a; Chaput et al. 2012).

The objective of these models was to reconstruct long term series (starting in the early 1970's) of abundance at sea before any marine fisheries (Pre Fishery Abundance, PFA, measured at the January 1 of the first winter spent at sea) and to forecast the returns of adult salmon to their natal rivers (homewaters). These models have been incorporated in a risk analysis framework to assess the consequences of mixed stock marine fisheries at West Greenland and Faroes on the returns (Friedland et al., 2005; ICES, 2015a) and to assess compliance of realized spawning escapement to conservation limits (biological references point below which the stock should not pass) at both the SU and CSG scales.

However, PFA models suffer from three weaknesses that hinder their relevance for analyzing the demographic processes driving the population's dynamics of European and American A. salmon populations and should be addressed in order to improve the scientific basis of A. salmon population assessment.

- First, PFA models used for formulating catch advice at ICES rely on a coarsely constructed stock-recruitment dynamic. Forecasts of the returns during the three years after the last assessment are based on forecasts of the productivity parameter defined as the productivity between a spawning potential (measure of the stock; expressed as a number of eggs potentially spawned each year for the two European CSG and as the potential number of spawners in the North American CSG) and abundance at the PFA stage (measure of the recruitment). This framework does not explicitly represent the population dynamics as a life cycle.

Statistical inferences on the time series of productivity parameters are susceptible to time series bias because the dynamic link between PFA and subsequent egg depositions is not represented (Massiot-Granier et al., 2014; Su and Peterman, 2012). Also, the lack of flexibility in the modelling framework also restricts the integration of the large amount of available data and knowledge on A. salmon demographics and population dynamics. As such, hypotheses on drivers and mechanisms of changes cannot be easily tested (Massiot-Granier et al., 2014).

- Second, the PFA modelling framework actually works as a combination of three models, what makes the workflow hard to handle. (1) A first model, the *run reconstruction* model relies on estimates of the abundance of fish returning to spawn and biological parameters (sex ratio, fecundity and mean proportion of smolts ages) to estimate the potential number of spawners or eggs (measure of the Stock) for each year of the time series. The same model is used to estimate the abundance of fish at the PFA stages (measure of the Recruitment), through a back calculation procedure (similar to a Virtual Population Analysis) using data on catches at sea and hypothesis on natural mortality rates at sea. Hence, the measures of the stock and the recruitment are derived from the same data, whilst they are considered independent in the rest of the process. (2) A second part of the modelling framework consists of estimating the productivity parameters between the Stock and the Recruitment for all years of the historical time series, and uses time series hypothesis (random walk) to forecast the evolution of the productivity parameter during three years after the last year of the assessment. (3) in a third phase, this forecast of the productivity parameters serves as a basis to forecast the PFA and the number of fish that returns to homewater based on catches scenarios at sea.
- Third and more importantly, different and independent PFA models were developed for the three CSG. Some core demographic hypotheses are not harmonized among these models. Specifically, the two European models explicitly consider 1SW and 2SW fish in the population dynamics, while the current model for NA, which was developed for catch advice purposes at West Greenland, only considers the dynamics of 2SW fish (Chaput et al., 2005). The NA model implicitly assumes that 2SW spawners only produce 2SW fish in future cohorts, and excludes contributions of 1SW and multi-sea-winter spawners. Temporal variations of productivities for NA SU consider only the 2SW component and are therefore not comparable to the PFA models built for the European CSG considering both 1SW and 2SW components in marine productivity. These structural differences in models preclude the simultaneous analysis of the population dynamics among all SU in the North Atlantic. This approach also ignores any covariance structure in the dynamics of the SU even though the SU may share common environments at sea and be jointly exploited in sea fisheries.

In this working paper, we develop a Bayesian life cycle modelling framework for the combined analysis of Atlantic salmon population dynamics across all SU in the North Atlantic Ocean. We extend the framework developed by Massiot-Granier et al. (2014) for one SU to include the dynamics of all SU of the three CSG (Northern Europe, Southern Europe and North America) within a single unified hierarchical Bayesian life cycle approach with populations following a similar life history process.

The model brought a major contribution to improve the scientific basis for Atlantic salmon stock assessment.

- It provides a framework for analyzing the mechanisms that shape population responses to variations in marine ecosystems. In particular, it allows for modelling covariations among all SU and for partitioning the effects of fisheries from the effects of environmental factors at a hierarchy of spatial scales, including at the level of the North Atlantic, of each CSG, and for each SU within a CSG.
- The integrated life cycle framework is also expandable and provides an opportunity to assimilate various sources of information to improve the ecological and biological realism of the model.
- Last, the life cycle model is a natural framework for forecasting population dynamics. The same model is used for both the inferences (hindcasting) and forecasting phases, and all the model properties are readily integrated into the forecast process. This model is a new important tool to provide catch options for any marine fisheries that operate on a mixture of stocks (e.g. the West Greenland salmon fishery) and can also be used to evaluate catch options for other fisheries, or to assess conservation measures for the different sea-age classes or the SU separately.

# 2 Outlines of the model used to fit the historical series of data

## 2.1 Model design

The life cycle model is formulated in a Bayesian hierarchical state-space framework (Buckland et al., 2004; Cressie et al. 2009; Parent & Rivot, 2012; Rivot et al., 2004) that incorporates stochasticity in population dynamics as well as observation errors. To keep the presentation concise, all model equations and data sources are detailed in Appendix 1.

### 2.1.1 Spatial structure

The model considers the dynamics of 24 SU (subscript $r = 1, \ldots, N$ with $N$=24) (Fig. 1):

- 6 SU from NA CSG, indexed by r = 1, …, 6: 1 = Newfoundland, 2 = Gulf, 3 = Scotia-Fundy, 4 = USA, 5 = Quebec and 6 = Labrador);
- 7 SU from the SE CSG, indexed by r = 7, …, 13: 7 = Ireland, 8 = UK (England and Wales), 9 = France, 10 = UK (Scotland east), 11 = UK(Scotland west), 12 = UK (Northern Ireland) and 13 = south-west Iceland);
- 11 SU from NE CSG, indexed by r= 14,…,24: 14=North-East Iceland, 15=Sweden, 16=South-East Norway, 17=South-West Norway, 18=Middle Norway, 19=North Norway, 20=Finland, 21=Russia Kola Barents, 22=Russia Kola White Sea, 23=Russia Arkhangelsk Karelia and 24=Russia River Pechora.

SU are defined on the basis of freshwater areas. All salmon populations within a SU are assumed to undertake a similar migration route at sea. Note that Germany and Spain (SE CSG) and SU from the Northern Europe CSG are not considered at this stage because of an incomplete time series of data.

### 2.1.2 Variability of life histories

The model is built in discrete time on a yearly basis (subscript $t = 1, \ldots, n$ with $n$=44 in this present application).

The population dynamic of each SU is represented by a homogeneous age- and stage-structured life cycle model, applied to all SU (Fig. 2). The model incorporates variations in the age of out-migrating juveniles from freshwater (i.e., smolt ages) and the sea-age of returning adults among SUs. Smolts migrate to sea after 1 to 6 years in freshwater (depending on SU). Following the approach used by ICES for catch advice purposes (ICES 2015a), only two sea-age classes are considered in the model: maiden salmon that return to homewaters to spawn after one year at sea, referred to as one-sea-winter (1SW) salmon, or grilse, and maiden salmon that return after two winters at sea (2SW). This is a simplification of the larger diversity of life history traits as some maiden fish may spend more than two winters at sea before returning to spawn, and some salmon return as repeat spawners. Maiden spawners older than 2SW are relatively rare in North America and Southern Europe and the six smolt-age by two sea-age combinations represent the essence of life history variation.

The model tracks the abundance of fish ($N_{s_{t,r}}$) for each SU (*r*) by year ($t$) and life stage ($s$), sequentially from eggs ($N_1$) to 1SW ($N_7$) or 2SW ($N_{10}$) spawners for the period considered (starting in 1971, year of return to rivers) (Fig. 2; Table 1). Spawners are fish that contribute to reproduction and that therefore survived all sources of natural and fishing mortality. The transition rates between stages $s$ for each SU (*r*) in year $t$ are denoted $\theta_{s_{t,r}}$.

### 2.1.3 Hypotheses to separate the sources of variability

As recognized by the data constraints already expressed in the existing PFA models used by ICES (ICES 2015a) and discussed by Massiot-Granier et al. (2014), the quality and information provided by the data are limited, which restricts the number of population dynamic parameters that can actually be estimated. The framework is primarily designed to estimate the abundance at various life stages along the life cycle, the exploitation rates of all fisheries, and the two parameters that implicitly assume that most of the temporal variability occurs during the first months of the marine phase: the post-smolt marine survival rates (from out-migrating smolts to the PFA stage as of January 1 of the first winter at sea) and the proportions of fish maturing as 1SW, for each SU. To separate the variability in the natural and fishing mortality during the freshwater and marine phase and in the proportion of fish that mature as 1SW, we use the framework described in ICES (2015a) and Massiot-Granier et al. (2014).

#### 2.1.3.1 Freshwater phase

The number of eggs spawned in each SU by year are derived from the annual number of returning 1SW and 2SW spawners and the SU specific sex-ratio and fecundity values, these are considered fixed and constant over time (Table 2).

In the absence of information on the total smolt production at the scale of SUs, the parameters of the freshwater phase (eggs to out-migrating smolt production) are fixed. In the baseline configuration presented here (but see Olmos et al. 2019 for a sensitivity analysis to other modelling option, including density dependence), the eggs-to-smolt survival is density-independent, and modelled as lognormaly distributed around a average of 0.007 (7 per mile) with random variations (CV=0.4) independent across SU and years. As fecundity and freshwater survival are fixed a priori, the only variation in the freshwater phase of the life cycle is due to these lognormal random deviations (no trends, no density dependence). This implicitly assumes that most of the changes in the stock productivity over time are the result of variations in dynamics in the marine phase.

The total number of smolts produced by a cohort is attributed to river-age classes using SU specific smolt age proportions which are considered fixed and constant over time (Table 2).

#### 2.1.3.2 Marine phase

Smolts of different ages migrating seaward in any year $t$ are pooled together once at sea (Fig. 2). Returns rates from smolts to 1SW and 2SW adults result from the combination of natural mortality, fishing mortality, and a maturation schedule. The PFA stage is defined as abundance of post-smolts at January 1 of the first winter at sea, and prior to any fisheries. Survival from smolts to the PFA stage is estimated and may vary among years and SUs (Table 3). Fish at the PFA stage can then mature (and return as 1SW adults) or delay maturation until the following winter (and return as 2SW adults). The proportion of fish maturing as 1SW is estimated and may vary with year and SUs (Table 3).

The natural mortality before the PFA stage is estimated, but the natural mortality rate after the PFA stage is fixed, assumed constant in time, homogeneous among all SUs, and identical for maturing and non-maturing fish ($M = 0.03 \cdot month^{-1}$; Table 1). Under this assumption, the proportion of the PFA abundance that matures is confounded with the mortality difference between 1SW and 2SW salmon (see Massiot-Granier et al. (2014) and Olmos et al. (2019) for a discussion).

Fishing mortality is modelled as a temporal sequence of fisheries operating on mixtures of SU along the migration routes, as well as on each SU in homewaters (Fig. 2; Tables 4 and 5). Fisheries exploitation rates are estimated. They may vary by year and SU and are assigned weakly informative priors (Tables 4 and 5).

### 2.1.4 Covariation among SUs

The model explicitly incorporates two components of temporal covariation among all SUs (Fig. 3). First, the post-smolt survival (denoted $\theta_{3_{t,r}}$) and the proportion of fish maturing as 1SW (denoted $\theta_{4_{t,r}}$) are modelled as multivariate random walks in the logit scale which captures spatial covariation associated with environmental stochasticity. Random variations are drawn from multivariate Normal distributions in the logit scale with variance-covariance matrices $\Sigma_{\theta_3}$ and $\Sigma_{\theta_4}$ (Minto et al., 2014; Ripa and Lundberg, 2000) (Table 3):

(1) $$\left(logit(\theta_{3_{t+1,r}})\right)_{r=1:N} \sim MVNormal\left(\left(logit(\theta_{3_{t,r}})\right)_{r=1:N}, \Sigma_{\theta_3}\right)$$

(2) $$\left(logit(\theta_{4_{t+1,r}})\right)_{r=1:N} \sim MVNormal\left(\left(logit(\theta_{4_{t,r}})\right)_{r=1:N}, \Sigma_{\theta_4}\right)$$

with N = the number of SU in the model (here N=24).

The pairwise correlation matrix $\rho$ can be calculated from the variance-covariance matrix:

$$(3) \qquad \rho = \sqrt{diag(\Sigma)}^{-1} \times \Sigma \times \sqrt{diag(\Sigma)}^{-1}$$

The second source of covariation among SU is the harvest dynamics of the sequential marine fisheries that operate on mixtures of SUs, with the portfolio of SU available for each fishery dependent on marine migration route hypotheses (Fig. 3).

## 2.2 Time series of data and likelihood

The model is fitted to time series of data for years $t = 1, \ldots, n$. It incorporates observation errors for the time series of returns and catches for each year and sea-age class separately. The full likelihood function for the general state-space model is built from the combination of all observation equations for the returns, homewater catches, and catches at sea, for 1SW and 2SW separately.

Building an integrated model (Maunder and Punt, 2013; Rivot et al., 2004; Schaub and Abadi, 2011) that explicitly integrates complicated observation models would dramatically increase the complexity of the full model. Therefore, a sequential approach (Michielsens et al., 2008; Staton et al., 2017) is used that consists of (*i*) processing observation models separately to reconstruct probability distributions that synthesize observation uncertainty around the time series of catches and returns for the 13 SUs; and (*ii*) using those distributions as likelihood approximations in the population dynamics state-space model.

Probability distributions for returns and catches are derived from a variety of raw data and observation models, specific to each SU (except for the mixed stock fisheries at sea) as originally developed by ICES to provide input for PFA models. These consist of:

- Time-series of estimates (approximated as logNormal distributions) of the number of maturing anadromous Atlantic salmon that return to homewaters for each of the 13 SU by 1SW and 2SW maiden sea-age classes. **In the application reported in** **this working paper, returns are directly derived from the Run Reconstruction models run by ICES WGNAS (a lognormal distribution was fitted to the Monte Carlo draws of estimated returns).** It is worth noting that those probability distributions are built in a separate step, independently from the life cycle model. In our new approach, the run reconstruction model is no more needed as such, as different methods than the ones used in the run reconstruction model (and each specific to each SU) could potentially be developed to reconstruct those probability distribution of returns.

- Time series of estimates (with observation errors, approximated as logNormal distributions) of homewater catches for each SU by sea-age class;
- Time series of estimates (with observation errors, approximated as logNormal distributions) of catches for the mixed stock fisheries at sea operating sequentially on combinations of SUs, and using additional data on the SU origin of the catches.

### 2.2.1 Data from ICES WGNAS 2015 were used in this working paper

The version of the model presented in this WP was fitted to the data of the ICES WGNAS report 2015 (ICES, 2015a). The time series of data is therefore 44 years from 1971 to 2014. Subscript $t = 1,..,n$ hence stand for the time series 1971 to 2014 with n=44.

ICES (2O15) provides a shorter time series of data for Northern NEAC SU because some data are missing for Norway for the first time of the time series before 1982. In order to have the same length of data series for all SU (1971-2014), the Norwegian data were complemented using some best guest hypotheses (*Com pers*. Geir Bolstad and Peder Fiske, NINA; see hereafter in the detailed description of the data).

### 2.2.2 Abundance of returns

Independent logNormal distributions were used to approximate the likelihood of the returns, described by means and coefficients of variation (CV) specific to the SU, year, and sea-age class (Fig. 4).

### 2.2.3 Homewater fisheries

The homewater fishery pools all fisheries capturing returning fish in coastal, estuarine and freshwater areas. Independent logNormal distributions were used to approximate the likelihood of the homewater catches, with means specific to each SU, year and sea-age class (Fig. 5). Because homewater catches are generally provided with small observation errors, we used logNormal distributions with relative errors arbitrarily fixed to a CV = 0.05 around the point estimates.

### 2.2.4 Distant marine fisheries

Catches of the distant marine fisheries are derived from the declared catches reported to ICES. Fish originating from North America and Europe have different migration routes at sea to eventually reach the common feeding grounds in West Greenland after the 1$^{st}$ winter at sea. The West Greenland (WG) fishery which potentially harvests non-maturing salmon from a mixture of stocks from all SU from North American and Europe (although the proportion of fish originating from Northern Europe is low; Fig. 1 & 3). The Faroes (Fa) fishery harvests non-maturing and maturing salmon from all SU of both European stock complexes (North and South). Other sea fisheries considered operate on a mixture of stocks from one CSG.

For each fishery considered, the likelihood equations associated with catches consist of logNormal distributions of observation errors of the total catches by sea age class, summed over all SU exploited by the fishery, combined with Dirichlet likelihood terms for the proportion of catches allocated to each SU when those data are available (Faroes and West Greenland fisheries). Observation errors on the total catches and on the proportions are considered independent across fisheries, years and SU.

Note that when data are used to allocate catches to the different SU, this may result in different SU being harvested non-homogeneously. For instance, the harvest rate estimated for a particular SU will be high if the proportion in the data used to allocate catches is higher than the proportion of this SU in the total abundance.

#### 2.2.4.1 Fisheries operating on a mixture of North American SUs

NA fish maturing in the first year at sea (1SWm) may be exploited on their return migrations to rivers in the marine fisheries of Newfoundland and Labrador (NFLD/LAB) and at Saint-Pierre et Miquelon (SPM) (Table 4). Salmon that do not mature during the first year at sea (1SWnm) may be caught in the LAB/NFLD marine fisheries and at WG as 1SWnm, and as 2SW salmon on their migration to home waters in the LAB/NFLD and SPM fisheries.

Catches of 1SWnm at WG may originate from any of the 24 SU from all CSG (Fig. 3). A compilation of individual assignment data based on discriminant analyses of scale characteristics and genetic analyses was used to allocate the catches in the WG fishery to the 24 SU (Bradbury et al. 2016a, 2016b; ICES 2017a; 2017b ; but see also Olmos et al. 2019 for more details) (Fig. 6).

LAB/NFLD and SPM fisheries exploit a mixture of SU from only NA (Fig. 3). Data and expert opinion are used to partition catches of 1SWm, 1SWnm, and 2SW in the LAB/NFLD fishery originating from Labrador from those originating from the other NA SU (ICES 2017a; 2017b)( Fig. 7). The SPM fishery is assumed to not catch any fish from Labrador and the exploitation rate of salmon from Labrador SU was fixed to zero in this fishery. Other than these assumptions and in the absence of data to differentially allocate catches to each of the six SU in NA, catches were assigned assuming that exploitation rates were homogeneous among the six SU (ICES 2017a; 2017b).

#### 2.2.4.2 Fisheries operating on mixtures of European SUs

1SWm fish from SE are susceptible to be harvested in the Faroes (FA) fishery before they return to homewaters (Fig. 3; Table 5). Fish that mature as 2SW may be first harvested at FA as 1SWnm in the first winter at sea, before migrating to the WG feeding grounds where they are susceptible to be harvested together with fish from NA. Those that survive the WG fishery are susceptible to be harvested at FA as 2SW fish before migrating back to their homewaters.

Total catches of 1SWm, 1SWnm, and 2SW at FA are allocated to each of the SU in SE and NE using limited genetic assignment data which are set as fixed and constant over time (ICES 2017a; Table 6; Fig. 8).

## 2.3 MCMC simulations, convergence and posterior checking

Bayesian posterior distributions were approximated using Monte Carlo Markov Chain (MCMC) methods in *Nimble* (https://r-nimble.org/) through the *rnimble* (www.Rproject.org) package.

Sampling efficiency for this model is relatively low, meaning that a long MCMC simulation is needed to obtain reasonable convergence to the posterior distribution and reliable results.

**We recommend the following MCMC configuration:**

- Use well chosen initial values for the MCMC chains. We recommend simulating initial values from the Nimble model to ensure the consistency of initial values with the model. We also recommend using initial values close to the posterior to avoid initializing the model in a region of the parameters space were the likelihood is too low. An R-code to simulate appropriate initial values for the MCMC chains is provided.
- Run at least two independent MCMC chains with dispersed initialization values. This is needed to check mixing.
- Run the model during a relatively long period before storing the results to let the algorithm adapt and optimize. We recommend to discard the first 10000 iterations before storing (burnin = 10000).
- Use at least 2500000 MCMC iterations after the burnin period for final inferences. In any case reduce the size of MCMC chains without carefully checking the convergence.
- Use a large thining of MCMC chains to avoid storing too long MCMC chains with poor information. The level of autocorrelation of MCMC chains is very high (still significant at lag 500) and we recommend a thinning of at least 500. Running 2500000 iterations with a thin=500 will result in a sample of 5000 iterations kept for inferences.

- Monitor mixing of the chains for all parameters, and formally assess convergence using the Gelman-Rubin statistic (Brooks and Gelman, 1998) as implemented in the R Coda package (gelman.diag()).

**Important note**:

The MCMC configuration above (2 chains in parallel; burnin = 10000; 2500000 iterations; thin = 500 → resulting in 5000 iterations/chain saved) takes ~ 72 hours (3 days) to run with a personal Laptop Intel Core i7 – 3.0Ghz). Work to reduce computational time is under progress.

# 3 Forecasting and risk analysis framework

Following ICES WGNAS practices, the life cycle model is used to forecast the population dynamics during five years starting after the last year of the assessment (in this application, forecast is therefore 2013-2017), based on different catches scenarios in the Faroes and Greenland mixed stock fisheries (Fig. 9). The forecasted abundance of returns after all marine distant fisheries (but before homewater fisheries) is then compared to the management objectives defined below.

The same life cycle model is used for fitting the historical time series and forecasting. An exception are the transitions that involve a fishing mortality, modelled by directly retrieving catches to the abundance (this is because scenarios are defined by fixing catches and not harvest rates). The post-smolt marine survival and the proportion maturing are forecasted following the multivariate random walks defined at equations (1)-(2). Because of the random walk hypothesis, the forecasted marine survival and proportion maturing during the forecasting period will remain at the same average level than the last year of the fitted time series, but with an uncertainty that increases with time due to error propagation through the random walk.

Parameters uncertainty is integrated using Monte Carlo simulation, by simulating multiple population trajectories with parameters randomly drawn in the posterior MCMC sample. In practice, forecasting uses a replicate of the life cycle model written in R so as the posterior MCMC samples can be used to quickly run multiple scenarios.

## 3.1 Management objectives- Conservations Limits (CLs)

Management objectives are based on Conservation Limits (CLs) as defined by ICES and NASCO. CLs are defined as the quantity of eggs that should be deposited by spawners to produce a desired production of smolts (Table 7). Following the principles adopted by NASCO (1998) CLs for North Atlantic salmon have been defined by ICES WGNAS as limit reference points, in the sense that having abundance of eggs spawned falls below these limits should be avoided with high probability.

Management objectives in SE and NE are to reach or exceed CLs for both 1SW and 2SW fish. However, in NA management objectives currently defined by ICES consider the 2SW fish component of the returns only. However, in this working paper, CLs for North America have been defined as the total required egg deposition for both sea age classes (1SW+2SW) for all SU including North America.

CLs used by ICES are only available at a more aggregated spatial scale than SU defined in our life cycle model (Table 7). Specifically, one CL is available for Scotland (sum of Eastern Scotland and Western Scotland in our model), one CL for Norway (sum of 4 SU in our model, South-East Norway, South-West Norway, Middle Norway and North Norway) and one CL for Russia (sum of 4 SU in our model, Russia Kola Barents Russia Kola White Sea, Russia Arkhangelsk Karelia and Russia River Pechora). To be compared to the CLs defined by ICES, returns of spawners in our model were then summed to match with the spatial scale considered for CLs.

## 3.2 Risk analysis framework for the Western Greenland and the Faroes fishery

We used probabilistic forecasts from the model to evaluate the probability that future returns of spawners fall below management objectives for different catch options in the Western Greenland and the Faroes fisheries.

SUs from NA, SE and NE are all potentially harvested by the West Greenland fishery (although the proportion of fish originating from Northern Europe is very low in West Greenland catches). A risk framework for the provision of catch advice for the West Greenland fishery has been applied since 2003 by NASCO and ICES (ICES, 2013). Only fish from SE and NE are potentially harvested at the Faroes fisheries. There is currently no agreed framework for the provision of catch advice for the Faroes fishery adopted by NASCO. However, NASCO has asked ICES, for a number of years, to provide catch options or alternative management advice with an assessment of risks relative to the objective of exceeding stock conservation limits for salmon in the European area (NE and SE complexes).

As an important contribution, our new life cycle model provide a unique framework for evaluating catch options for the Faroes and West Greenland separately or simultaneously and for all SU separately or simultaneously.

For the purpose of demonstration, in this working paper 36 scenarios that consider both Faroes and West-Greenland catches were then built by crossing 6 combinations of catches from 0 to 250 tons (0, 50, 100, 150, 200, 250) for both the Western Greenland and Faroes fisheries. For each scenario, catches options were converted to number of fish really caught using mean weight of fish following ICES (2015a). Population dynamics was simulated with homewater catches and proportions to allocate catches at West Greenland and Faroes fisheries to the different SU fixed to the average of the last five years of the time series of data (2008-2012), and 0 catches for other distant fisheries.

For the West Greenland fishery, the catch of 1SW salmon of North American and European origins is further discounted by the fixed sharing fraction (Fna) historically used in the negotiations of the West Greenland fishery, that is a 40%:60% West Greenland:(North America & Europe) split. For instance, in a scenario with a 100t quotas, a total of 250t are actually caught, 150t are reserved for the Western Greenland fishery and 100t are reserved for the North American and European commercial fishery (note that the scientific advice given by ICES since several years is a quotas of 0t).

For each scenario, we provide forecasts during five years (in this application, 2013-2017) starting after the last year of our assessment model (2012). Monte Carlo simulations are run to integrate over both process' errors and parameters' uncertainty. Parameters uncertainty is integrated by randomly sampling the parameters in the joint Bayesian posterior distribution probability around parameters, which captures the covariance structure among the parameters. For a given set of parameters, the population dynamics is simulated including process error (i.e., inter-annual variability).

The probability of each SU (or aggregation of SU as defined in Table 7) achieving its CL individually and the probability of this being achieved by all management units simultaneously within a same CSG (i.e. in the same given year) are calculated from Monte Carlo trials. This allows managers to evaluate both individual and simultaneous achievement of management objectives in making their management decisions.

# 4 Results - Fitting the life cycle model to the historical (1971-2014) time series of data

Those results are derived from Maxime Olmos PhD (Olmos, 2019). They were obtained using the data from ICES 2015.

## 4.1 A widespread decline of abundances in all CSG

Posterior estimates of returns (total 1SW + 2SW; Fig. 10a) show consistent declining trends from the early 1970s to the 2010s in all CSG. Returns at the end of the time series were estimated to be ~50% of the abundances at the beginning of the 1970s for NA and SE CSG and ~30% for NE CSG. In NA CSG returns show an increase in abundance from 2003 (mostly due to an increase in Labrador and Newfoundland) that is not observed for the two other CSG.

The average proportion of 1SW fish in returns is different between the three CSG (Fig. 10b). The proportion of 1SW in returns is lower in the NE CSG, which is characterized by a high proportion of fish spending more than one winter at sea. The Southern European CSG has the highest average proportion of 1SW in returns. The three CSG exhibit similar temporal trends in the proportions of 1SW salmon in returns (Fig. 10b). The average time trend shows a consistent increasing trend from the early 1970s to the early 1980s, followed by a plateau or even a slight decline for the NE CSG.

Trends in spawner and return abundances may differ due to variations in homewater fishery exploitation rates (Fig. 11). Egg depositions follow the same general temporal trends as spawners (Fig. 11e). The proportion of eggs spawned by 1SW is highly variable between the three CSG (Fig. 11f). Contrast between the three CSG corresponds to the contrast in the proportion of 1SW in the return augmented by the difference in the average number of eggs spawned per fish that is particularly high for 2SW fish in NE (because of higher female-biased sex ratio and higher average size of fish in NE).

Time series of total PFA in each CSG show very similar continuous declines by a factor 3, between the 1970s and the 2010s (Fig. 11) with a stronger decline for the NA CSG. The decline in PFA is marked by a strong decrease in abundances in the 1990s.

## 4.2 Coherence in temporal variations of post-smolt survival and proportion of fish maturing as 1SW

### 4.2.1 Post-smolt survival rate

The time-series of post-smolt survival for the 24 SU show a common decreasing trend over years (Fig. 12). The trends averaged over all SU of the same CSG exhibit slightly different tendencies over the years. Those patterns are consistent with the decline observed in the abundance at the PFA stage. The post-smolt survival in NA exhibit a strong decline by a factor 3 in the period 1985-1995. This decline is also observable in SE with a sharp decline by a factor 1.8 in 1987. The sharp decline in the late 80's-early 90's is less visible in NE. Trend in NE shows a continuous and smoothed decline over the period.

The majority of pairwise correlations are positive, with a median correlation among all SU of 0.084 ± 0.139 (correlations are calculated in the logit scale; Fig. 13). In general, correlations are stronger between geographically close SU. The results show strong correlations for SU within NA (0.333), followed by SE (0.138) and NE (0.083). Correlations between the NE SU are stronger for the block of SU going from Sweden (East) to Russia-KB (West). Covariance and correlation in the temporal variations of the probability to mature as 1SW

### 4.2.2 Proportion of fish maturing as 1SW

Time trends in the proportion of fish that mature as 1SW also show a strong coherence among SU. These are in accordance with the expectation of higher correlations between SU of the same CSG.

Overall, there is an increasing trend from the 1970s to the 1990s that corresponds to declines in the proportions of 2SW fish in the returns followed by a levelling off or even a decline from the 2000s (Fig. 14).

All time trends are consistent with the average trend, except for France which shows a consistent decline during the entire period. Consistently with the low proportion of 1SW observed in the returns, the two most eastern SU, Russia-AK and Russia-RP, and US differ from the others SU with a very low probability of maturing.

As observed for the post-smolt survival, most of the pairwise correlations are positive across the 24 SU, with an average correlation of 0.1 (correlations are calculated in the logit scale; Fig. 15). In general, the correlations are stronger for geographically close SU. The results show strong correlations for SU within NA (0.409), followed by SE (0.149) and NE (0.087).

# 5 Results - Forecasting and risk analysis

## 5.1 Eggs deposition compared to CLs

The model allows for forecasting abundances for all life stages in the model.

As an example of forecasts results, the abundance of eggs (sum 1SW and 2SW) deposited by spawners in the 17 management units (SU or aggregation of SU) obtained under the scenarios of 0 catches in both Faroes and West Greenland fisheries can be compared to the CLs (Fig. 16). Results show how uncertainty in the forecasts increases with forecasting horizon. This is mostly the consequence of uncertainty propagation through time in forecasts of the post-smolts survival and proportion maturing modelled as multivariate random walks. Evaluating catch options for mixed stock marine fisheries.

The probability that the eggs deposition achieve the CLs under any fishing scenarios is directly quantified through Monte Carlo draws (see next section).

### 5.1.1 Catch option for the West Greenland mixed stock fishery (0 catches at Faroes)

The probabilities of achieving management objectives are higher for the stocks in Northern and Southern Europe (Fig. 17). Stocks from Northern Europe have the highest probabilities of achieving their management objectives (probabilities between 0.5 and 1 for the no fishery scenario). In Southern Europe, Northern Ireland, Southwest Iceland, Scotland and England and Wales have the highest probabilities of achieving their management objectives (probabilities between 0.4 and 1 for the no fishery scenario). In contrast, Ireland and France from SE, and stocks from NA such as US, Scotia-Fundy and Gulf have very low probabilities (between 0 and 0.6 for the no fishery scenario) of achieving their management objectives. As expected, different catch options at West Greenland have minimal influence on the probability of achieving management objectives for stocks that represent only a very low proportion of the catches at West Greenland, such as all stocks of NE (that represent less than 5% of the total fish harvested in West Greenland) and most of the stocks of SE. Because they present the highest exploitation rate at WG, stocks from NA such as Labrador, Quebec, and Gulf have their probability of achieving management objectives decreasing when catch options increase.

Scenarios of catches at WG (including zero catches) provide a null, a very low and a small probability of simultaneously achieving the management objectives for all stock units (or aggregated stock units) from NA, SE and NE, respectively (Fig. 17).

### 5.1.2 The Faroes mixed stock fisheries (O catches at West Greenland)

Southwest Iceland, Northeast Iceland, England and Wales, Norway and Russia have the highest probabilities of achieving their management objectives (probabilities between 0.6 and 1 for the no fishery scenario) (Fig. 18). By contrast, Ireland, Northern Ireland and France have low probabilities (between 0 and 0.8 for the no fishery scenario) of achieving their management objectives. As expected, different catch options at Faroes have influence on the probability of achieving management objectives, except for Ireland and Northern Ireland that represent only a very low proportion of catches at Faroes and for Iceland where the return are always well above CLs for all scenarios).

Scenarios of catches at Faroes (including zero catches) provide a quasi-null probability of simultaneously achieving the management objectives for all stocks of SE (Fig. 18). However, for stock units (or aggregated stock units) from NE, the probability of simultaneously achieving the management objectives is between 5% and 80%.

### 5.1.3 Evaluating catch options for West Greenland and Faroes fisheries simultaneously

The new life cycle model allows for evaluating simultaneously catch options in Faroes and West Greenland. Because less than 4% of fish from NE are harvested in West Greenland, and because no fish from NA move to Faroes, we only report results for stocks from SE (Fig. 19).

As already shown with the independent assessment of Faroes and West Greenland fisheries, Southwest Iceland have the highest probabilities of achieving their management objectives and Ireland the lowest.

Interestingly, eggs deposition in France, England&Wales and Scotland are more sensitive to catch options in the Faroes fisheries than in the West Greenland fishery. Indeed, those three stock units (or aggregated stock units) represent 50% of the catches in the 1SW maturing Faroes fishery, and a few amount of fish harvested in the West Greenland fishery.

# Figures and Tables

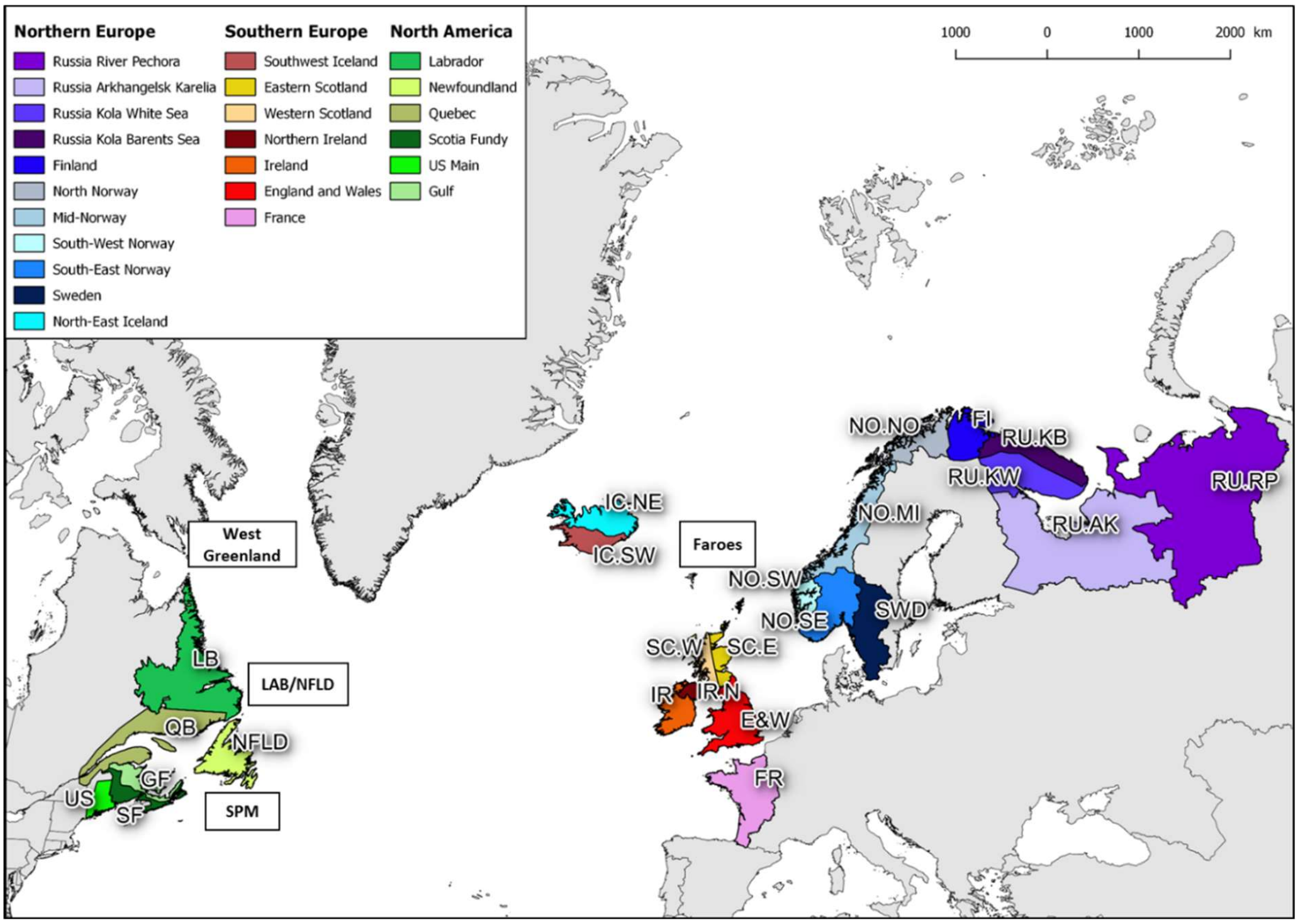

**Figure 1**. The 24 stock units considered in North Atlantic. Stock units of North America: NFDL=Newfoundland, GF=Gulf, SF=Scotia-Fundy, US=USA, QB=Quebec and LB=Labrador ; Stock units in Southern Europe: IR=Ireland, E&W=England&Wales, FR=France, E.SC=Eastern Scotland, W.SC=Western Scotland, N.IR=Northern Ireland, IC.SW=South-West Iceland ; Stocks units in Northern Europe: FI=Finland, IC.NE=North-East Iceland, NO.MI=Middle Norway, NO.NO=North Norway, NO.SE=South-East Norway, NO.SW=South-West Norway, RU.AK=Russia Arkhangelsk Karelia, RU.KB=Russia Kola Barents Sea, RU.KW=Russia Kola White Sea, RU.RP=River Pechora, SWD=Sweden. Germany and Spain are not included in the model. Boxes indicate the main fisheries at sea operating on mixed stocks: Faroes, West Greenland, Labrador and Newfoundland (LAB/NFLD), and Saint Pierre and Miquelon (SPM).

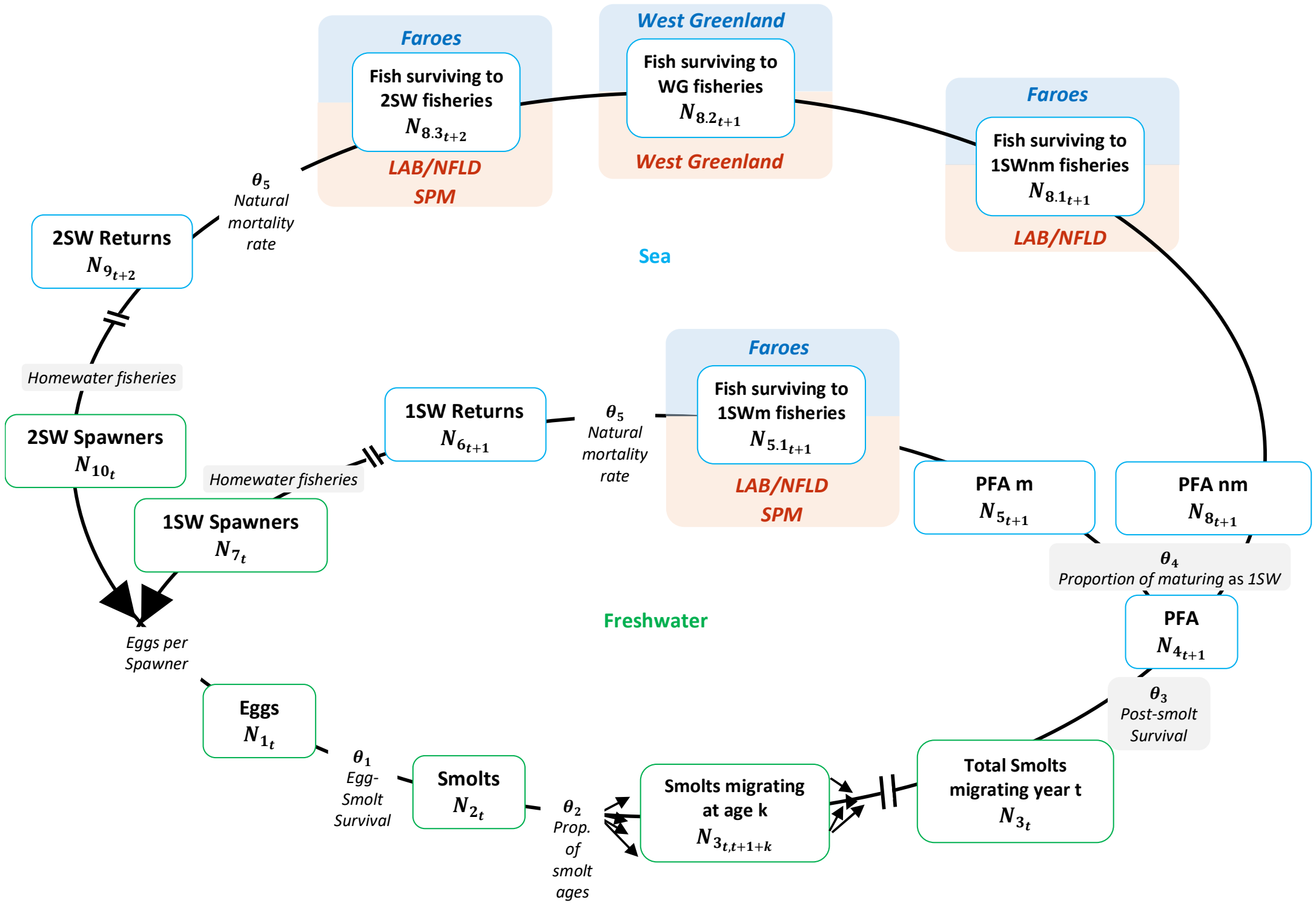

**Figure 2**. Structure of the age- and stage-based life cycle model. N1,t,r is the total number of eggs calculated from N7,t,r and N10,t,r. N3,t,r is the total number of smolts migrating in year t, as the sum of all smolts of age a=1,…,6 that migrate at year t. Red and blue boxes represent the migration routes with the associated sequential fisheries at sea that are specific for SU from NA and SE, respectively. Double bars indicate where cut in the time indices have been introduced to make notations easier. Light-shaded stages (eggs per spawner, proportion of smolt ages, and natural mortality) are transitions with parameters fixed or assigned with very informative prior distributions. Shaded (dark, blue or red) stages (exploitation rates, post-smolt survival, and proportion of fish maturing as 1SW) are parameters estimated from time series of data.

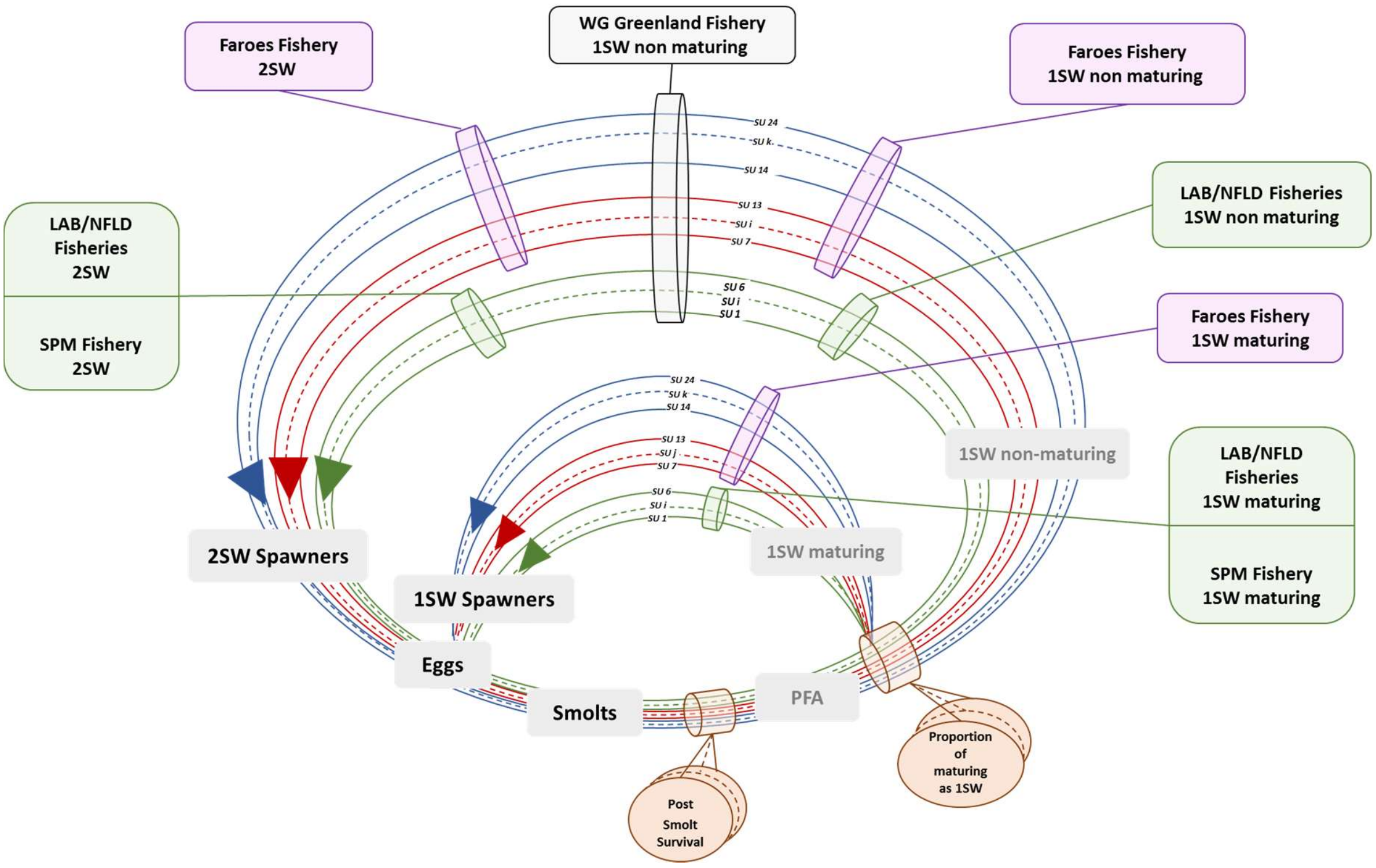

**Figure 3.** Structure of the age- and stage-based life cycle model for the 24 SU. Sources of covariation are two-fold: 1) covariations in the time series of post-smolt survival and proportion maturing as 1SW; 2) covariations through fisheries operating on mixtures of SU at sea. Grey boxes: different stages during the marine (grey) and freshwater (black) phases. Green circles belong to NA SU, Red circles belong to SE SU and blue circles belong to NE SU. Cylinder: sources of covariations among the 24 SU. Orange cylinders: key parameters (post-smolt survival and maturing probability). Purple cylinders: fisheries operating on mixture of SE and NE SU. Green cylinders: fisheries operating on mixture of NA SU. Grey cylinders: fisheries operating on mixture of NA, SE and NE SU.

**Table 1.** Summary of the main life stages and transitions of the life cycle model.

| Stage | Transition | Parameters | | Observation Equation |
|---|---|---|---|---|
| **N1**: Eggs | $N7_t \rightarrow N1_t$<br>$N10_t \rightarrow N1_t$ | *Sex ratio (SR)*<br>*Fecundity* (F) | Fixed<br>Fixed | No |
| **N2**: Total number of Smolt | $N1_t \rightarrow N2_t$ | *Freshwater survival* $(\theta_1)$ | Fixed | No |
| **N3**: Number of smolts in each age classe (6 age classes) | $N2_t \rightarrow$ $N3_{t+1+1}$, $N3_{t+1+a}$, $N3_{t+1+6}$ | *Proportion of smolt age* (Psmolt) | Estimated (Informative prior) | No |
| **N3tot**: Total number of smolts migration year t | $N3_{tot_t}$ | | | |
| **N4** : PFA<br>Pre Fishery Abundance | $N3_{tot_t} \rightarrow N4_{t+1}$ | *Post−smolt survival* (θ3) | Estimated (Multivariate random walk with covariation among SU) | No |
| **N5** : PFA maturing | $N4_t \rightarrow N5_t$ | *Maturing probability* (θ4) | Estimated (Multivariate random walk with covariation among SU) | No |
| **N8** : PFA non maturing | $N4_t \rightarrow N8_t$ | | | |
| **N5.1** 1SW maturing ($1SWm$) Faroes fishery | $N5_t \rightarrow N5.1_t$ | *Natural mortality* $(M)$<br>*Harvest rate* | Fixed<br>Estimated (uninformative prior) | Catches Faroes $1SWm$ observed with LogNormal errors and known variance |
| **N8.1** : 1SW non maturing ($1SWnm$) Faroes fishery | $N8_t \rightarrow N8.1_t$ | *Natural mortality* $(M)$<br>*Harvest rate* | Fixed<br>Estimated (uninformative prior) | Catches Faroes $1SWnm$ observed with LogNormal errors and known variance |
| **N8.2** : $2SW$ Faroes fisheries | $N8.1_t \rightarrow N8.2_{t+1}$ | *Natural mortality* $(M)$<br>*Harvest rate* | Fixed<br>Estimated (uninformative prior) | Catches Faroes $2SW$ observed with LogNormal errors and known variance |
| **N6** : Returns $1SW$ | $N5.1_t \rightarrow N6_t$ | *Natural mortality* $(m)$ | Fixed | Returns $1SW$ observed with LogNormal errors and known variance |
| **N9** : Returns $2SW$ | $N8.2_t \rightarrow N9_t$ | *Natural mortality* $(m)$ | Fixed | Returns $2SW$ observed with LogNormal errors and known variance |
| **N7** : Spawners $1SW$ | $N6_t \rightarrow N7_t$ | *Harvest rate* | Estimated (uninformative prior) | Catches observed with LogNormal errors and fixed variance |
| **N10** : Spawners $2SW$ | $N9_t \rightarrow N10_t$ | *Harvest rate* | Estimated (uninformative prior) | Catches observed with LogNormal errors and fixed variance |

**Table 2**. Parameters fixed or drawn in tight informative priors for the 24 stock units (Source: ICES 2015a). Note that the number of eggs per fish includes the proportion of females in spawners.

| | | NAC | | | | | | S.NEAC | | | | | | |
|---|---|---|---|---|---|---|---|---|---|---|---|---|---|---|
| | | **LB** | **NF** | **QB** | **GF** | **SF** | **US** | **FR** | **E&W** | **IR** | **N.IR** | **SC.W** | **SC.E** | **IC.SW** |
| **Egg to smolts survival** | | $\theta_{1_{t,r}}$ ~ Lognormaly distributed with average value $\mathbb{E}_{\theta_1} = 0.007$ and inter-annual variability $CV_{\theta 1} = 0.4$ | | | | | | | | | | | | |
| **Proportion of smolt ages** | $\boldsymbol{psm_{1,r}}$ | 0 | 0 | 0 | 0 | 0 | 0.377 | 0.917 | 0.23 | 0.05 | 0.38 | 0.2 | 0.05 | 0 |
| | $\boldsymbol{psm_{2,r}}$ | 0 | 0.041 | 0.058 | 0.398 | 0.6 | 0.52 | 0.083 | 0.75 | 0.75 | 0.59 | 0.5 | 0.45 | 0.05 |
| | $\boldsymbol{psm_{3,r}}$ | 0.077 | 0.598 | 0.464 | 0.573 | 0.394 | 0.103 | 0 | 0.02 | 0.2 | 0.03 | 0.3 | 0.45 | 0.73 |
| | $\boldsymbol{psm_{4,r}}$ | 0.542 | 0.324 | 0.378 | 0.029 | 0.006 | 0 | 0 | 0 | 0 | 0 | 0 | 0.05 | 0.21 |
| | $\boldsymbol{psm_{5,r}}$ | 0.341 | 0.038 | 0.089 | 0 | 0 | 0 | 0 | 0 | 0 | 0 | 0 | 0 | 0 |
| | $\boldsymbol{psm_{6,r}}$ | 0.04 | 0 | 0.01 | 0 | 0 | 0 | 0 | 0 | 0 | 0 | 0 | 0 | 0 |
| **Natural mortality rate (per month) after the PFA stage (for 1SW and 2SW fish)** | | $M = 0.03 \cdot month^{-1}$ | | | | | | | | | | | | |
| **Migration duration between stages** | | See Table 4 and 5 | | | | | | | | | | | | |
| **Number of eggs per fish** | $\boldsymbol{eggs_{1,r}}$ | 1500 | 3000 | 468 | 547 | 917 | 200 | 1552 | 1350 | 2040 | 1972 | 2000 | 2000 | 2501 |
| | $\boldsymbol{eggs_{2,r}}$ | 5500 | 4000 | 6402 | 5956 | 6107 | 5500 | 5520 | 4550 | 5950 | 4069 | 6000 | 6000 | 6149 |

**Table 2**. (continuing)

| | | N.NEAC | | | | | | | | | | |
|---|---|---|---|---|---|---|---|---|---|---|---|---|
| | | IC.NE | SW | NO.SE | NO.SW | NO.MI | NO.NO | FI | RU.KB | RU.KW | RU.AK | RU.RP |
| **Egg to smolts survival** | | $\theta_{1_{r,t}}$ ~ Lognormaly distributed with average value $\mathbb{E}_{\theta_1} = 0.007$ and interannual variability $CV_{\theta 1} = 0.4$ | | | | | | | | | | |
| **Proportion of smolt ages** | $\boldsymbol{psm_{1,r}}$ | 0 | 0.07 | 0 | 0 | 0 | 0 | 0 | 0 | 0 | 0 | 0 |
| | $\boldsymbol{psm_{2,r}}$ | 0.09 | 0.65 | 0.379 | 0.379 | 0.057 | 0.003 | 0 | 0.05 | 0.1 | 0.05 | 0 |
| | $\boldsymbol{psm_{3,r}}$ | 0.37 | 0.25 | 0.524 | 0.524 | 0.608 | 0.263 | 0.26 | 0.4 | 0.6 | 0.55 | 0.6 |
| | $\boldsymbol{psm_{4,r}}$ | 0.49 | 0.03 | 0.094 | 0.094 | 0.316 | 0.583 | 0.59 | 0.4 | 0.3 | 0.4 | 0.4 |
| | $\boldsymbol{psm_{5,r}}$ | 0.05 | 0 | 0.004 | 0.004 | 0.019 | 0.138 | 0.14 | 0.1 | 0 | 0 | 0 |
| | $\boldsymbol{psm_{6,r}}$ | 0 | 0 | 0 | 0 | 0 | 0.012 | 0.01 | 0.05 | 0 | 0 | 0 |
| **Natural mortality rate (per month) after the PFA stage (for 1SW and 2SW fish)** | | $M = 0.03 \cdot month^{-1}$ | | | | | | | | | | |
| **Migration duration between stages** | | See Table 4 and 5 | | | | | | | | | | |
| **Number of eggs per fish** | $\boldsymbol{eggs_{1,r}}$ | 1974 | 1500 | 887 | 887 | 1050 | 450 | 600 | 350 | 2700 | 450 | 450 |
| | $\boldsymbol{eggs_{2,r}}$ | 7350 | 4200 | 4944 | 4944 | 5128 | 6673 | 10010 | 10000 | 4200 | 9600 | 10500 |

**Table 3**. Parameters of the marine phase drawn in non-informative prior and for which update from the data is expected. Note that all those parameters concern the marine phase of the life cycle. All parameters of the freshwater phase are considered known or drawn in very tight informative prior distribution.

| | |
|---|---|
| Non diagonal (plain) N×N variance-covariance matrix (N=24)<br><br>Two different matrix for the post-smolt survival ($\Sigma_{\theta_3}$) and for the proportion of fish maturing as 1SW ($\Sigma_{\theta_4}$) | $\Sigma_\theta = \begin{pmatrix} {\sigma^2}_{\theta_{1,1}} & \dots & {\sigma^2}_{\theta_{1,N}} \\ \dots & \dots & \dots \\ {\sigma^2}_{\theta_{N,1}} & \dots & {\sigma^2}_{\theta_{N,N}} \end{pmatrix}$<br>${\Sigma_\theta}^{-1}$ ~Wishart($\Omega, \delta$) with scale matrix $\Omega$ set as the N×N identity matrix and $\delta$ the degree of freedom set to N |
| Exploitation rate of all fisheries $f$ (marine and freshwater) for any year $t$ and stock unit $r$ | $h_{f_{t,r}} \sim Beta(1,2)$ |

**Table 4**. Summary of the duration among stages and the sequential fisheries (operating on mixed stocks at sea and homewater fisheries) for stock units in the North American continental stock grouping (Source: ICES 2015a, Prévost et al., 2009).

| North American continental stock grouping | | |
|---|---|---|
| **Stages/Fisheries** | **Migration duration** | **Exploitation rate** |
| PFA maturing | | |
| ↓ | 7 months | |
| 1SWm NFDL/LB Fisheries | | Variable among years<br>Homogeneous among SU |
| ↓ | 0.5 months | |
| 1SWm SPM Fishery | | Variable among years<br>Homogeneous among SU |
| ↓ | 0.5 months | |
| Returns 1SW | | |
| ↓ | 0 | |
| 1SW homewater Fishery | | Variable among SU |
| ↓ | 0 | |
| Spawners 1SW | | |
| PFA non maturing | | |
| ↓ | 7 months | |
| 1SWnm NFDL/LB Fisheries | | Variable among years<br>Homogeneous among SU |
| ↓ | 2 months | |
| 1SWnm West Greenland Fishery | | Variable among years and SU<br>+ data to allocate catches among SU |
| ↓ | 3 months | |
| 2SWm NFDL/LB Fisheries | | Variable among years<br>Homogeneous among SU |
| ↓ | 5 months | |
| 1SWm SPM Fishery | | Variable among years years<br>Homogeneous among SU |
| ↓ | 0.5 months | |
| Returns 2SW | | |
| ↓ | 0 | |
| 2SW homewater Fishery | | Variable among years and SU |
| ↓ | 0 | |
| Spawners 2SW | | |

**Table 5**. Summary of the duration among stages and the sequential fisheries (operating on mixed stocks at sea and homewater fisheries) for stock units in the Southern and Northern European continental stock grouping (Source: ICES 2015a, Potter, 2016).

| Southern Europe continental stock grouping | | |
|---|---|---|
| **Stages/Fisheries** | **Migration duration** | **Exploitation rate** |
| PFA maturing | | |
| ↓ | 0.5 months | |
| 1SWm Faroes Fishery | | Variable among years and SU + data to allocate catches among SU |
| ↓ | 7.5 months | |
| Returns 1SW | | Variable among years and SU + data to allocate catches among SU |
| ↓ | 0 | |
| 1SW homewater Fishery | | Variable among years and SU |
| ↓ | 0 | |
| Spawners 1SW | | |
| PFA non maturing | | |
| ↓ | 0.5 months | |
| 1SWnm Faroes Fishery | | Variable among years and SU + data to allocate catches among SU |
| ↓ | 8.5 months | |
| 1SWnm West Greenland Fishery | | Variable among years and SU + data to allocate catches among SU |
| ↓ | 5 months | |
| 2SWm Faroes Fishery | | Variable among years and SU + data to allocate catches among SU |
| ↓ | 3.5 months | |
| Returns 2SW | | |
| ↓ | 0 | |
| 2SW homewater Fishery | | Variable among years and SU |
| ↓ | 0 | |
| Spawners 2SW | | |

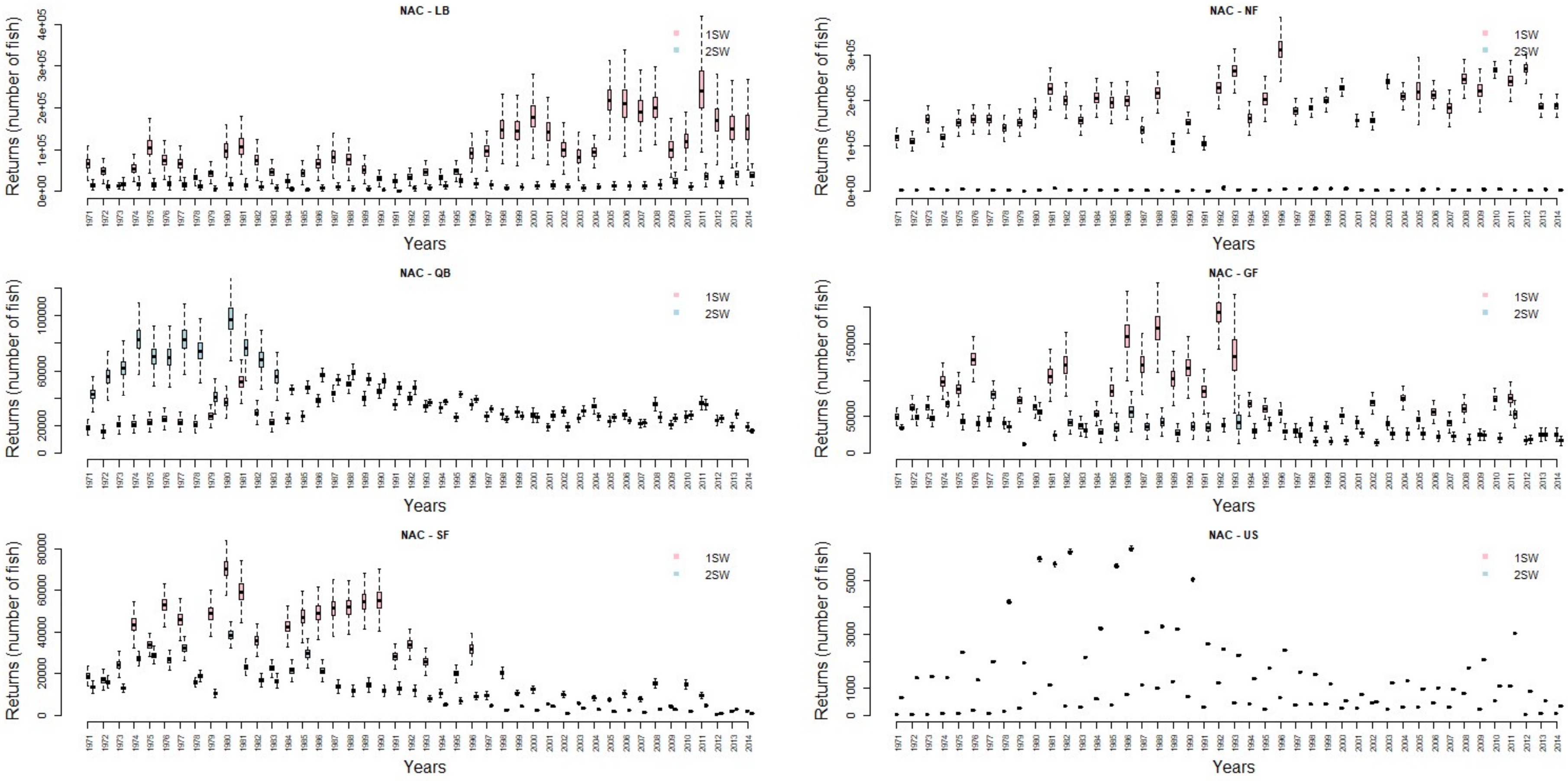

**Figure 4**. Probability distributions of the number of fish returning as 1SW (white boxplots) and 2SW (grey boxplots) in each SU of North America (Source: ICES 2015a).

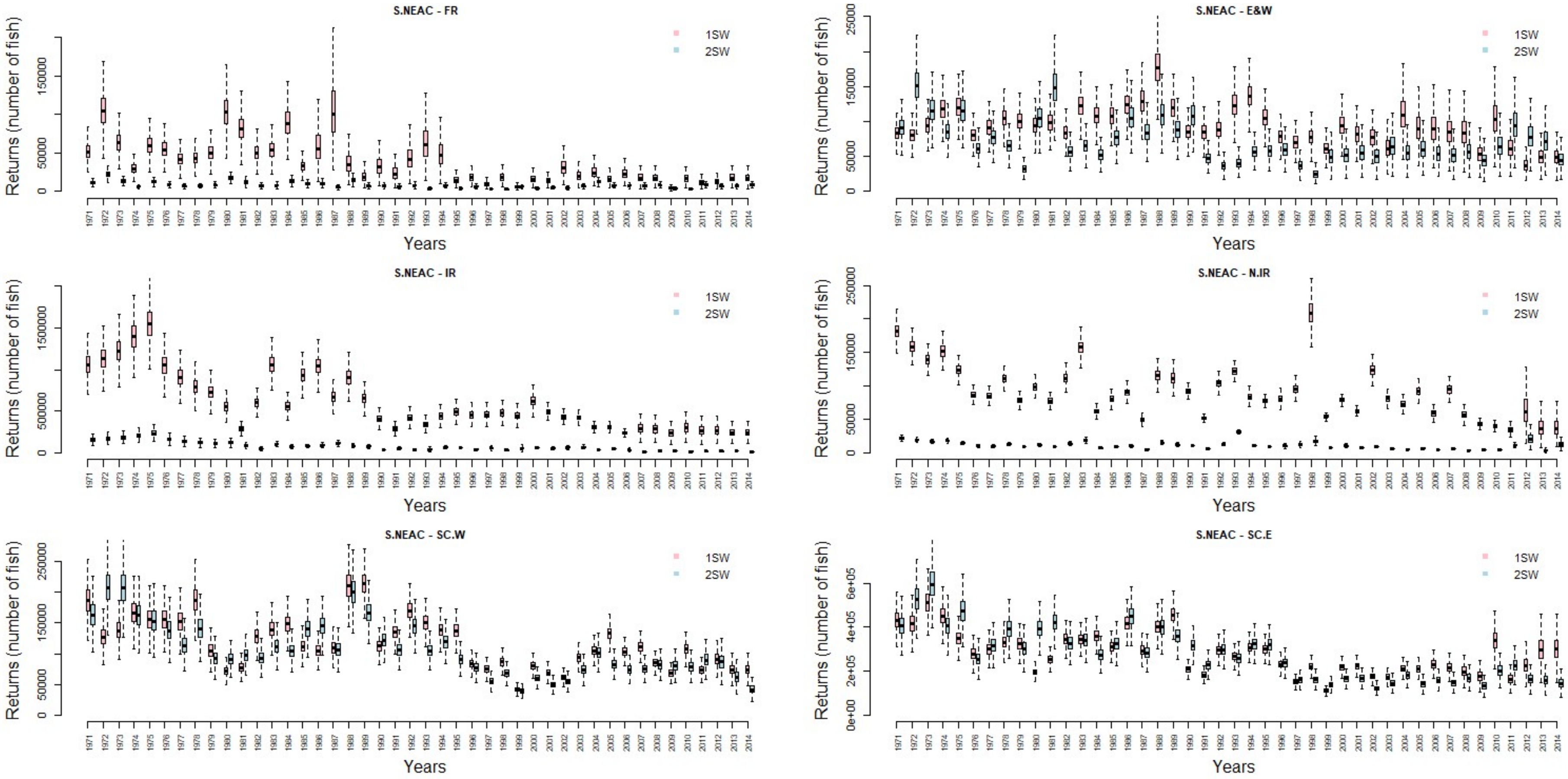

**Figure 4** (Continuing). Probability distributions of the number of fish returning as 1SW (white boxplots) and 2SW (grey boxplots) in each SU of Southern Europe (Source: ICES 2015a).

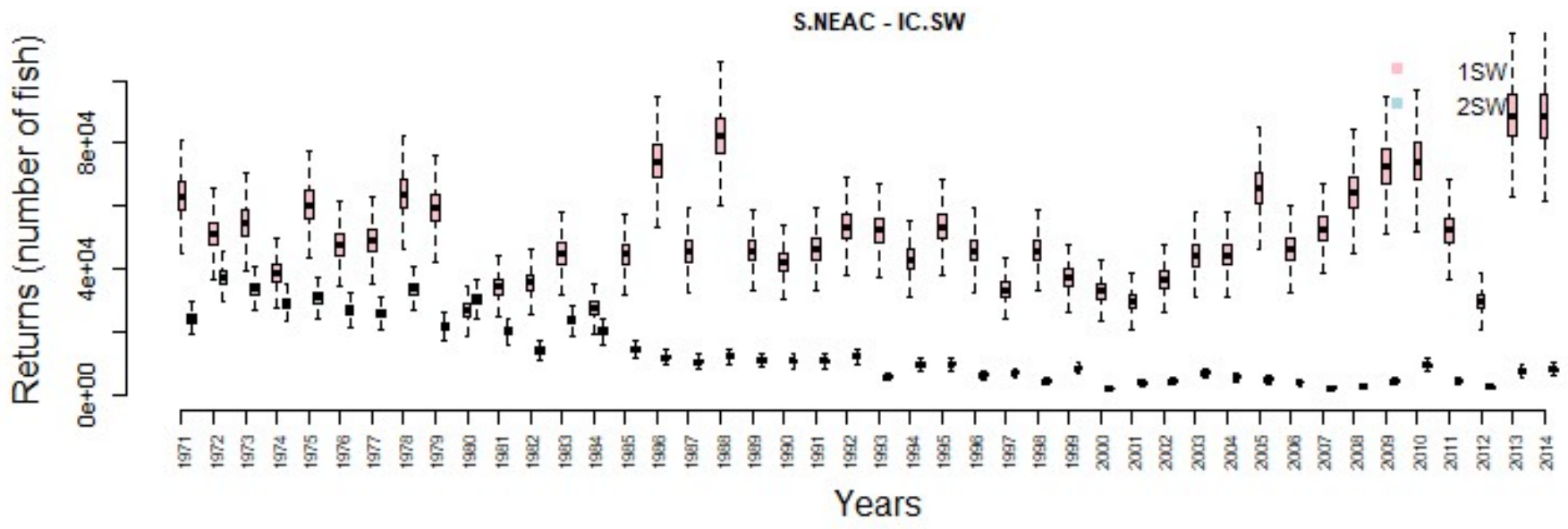

**Figure 4** (Continuing). Probability distributions of the number of fish returning as 1SW (white boxplots) and 2SW (grey boxplots) in each SU of Southern Europe (Source: ICES 2015a).

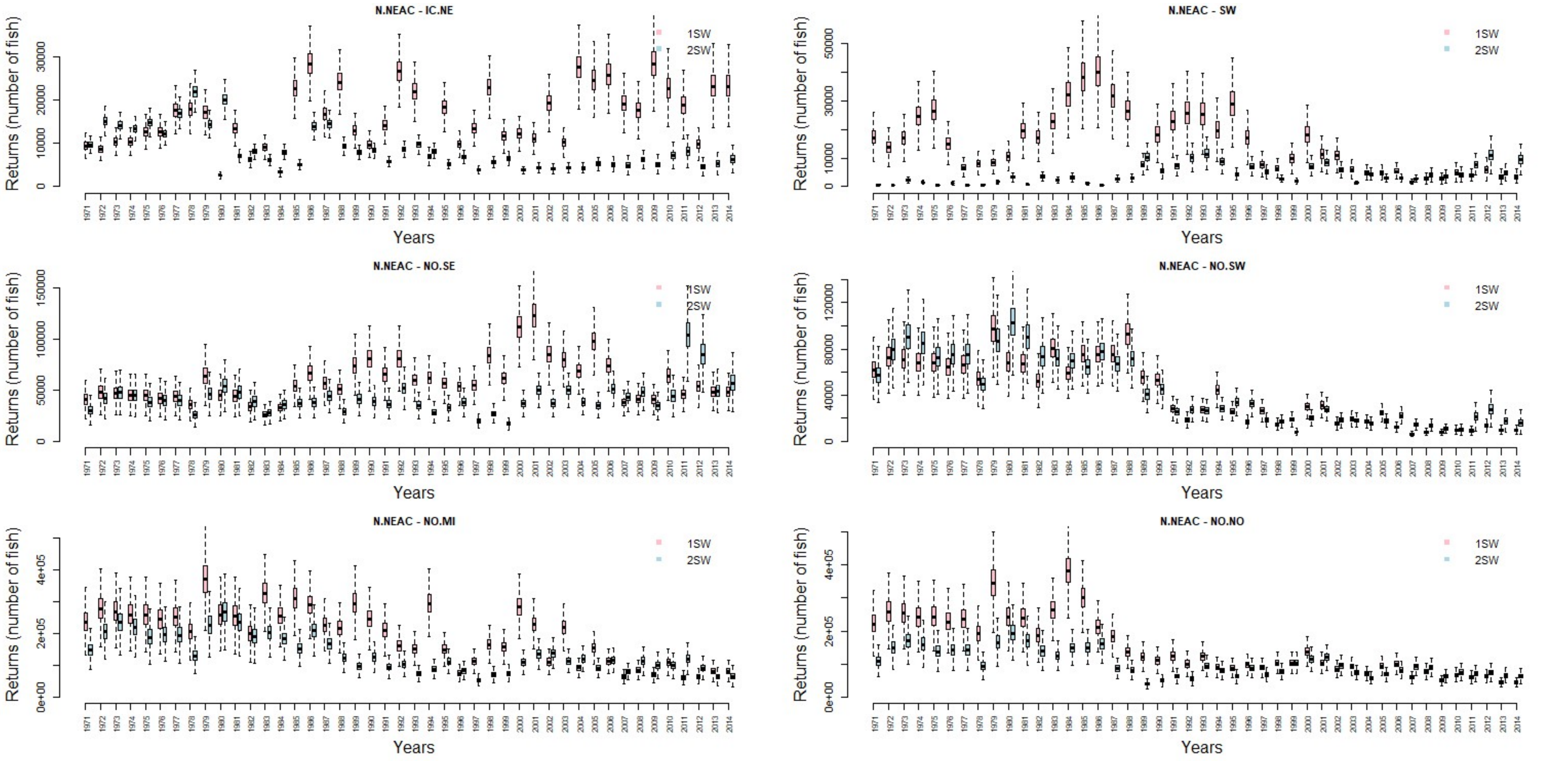

**Figure 4** (Continuing). Probability distributions of the number of fish returning as 1SW (white boxplots) and 2SW (grey boxplots) in each SU of Northern Europe (Source: ICES 2015a).

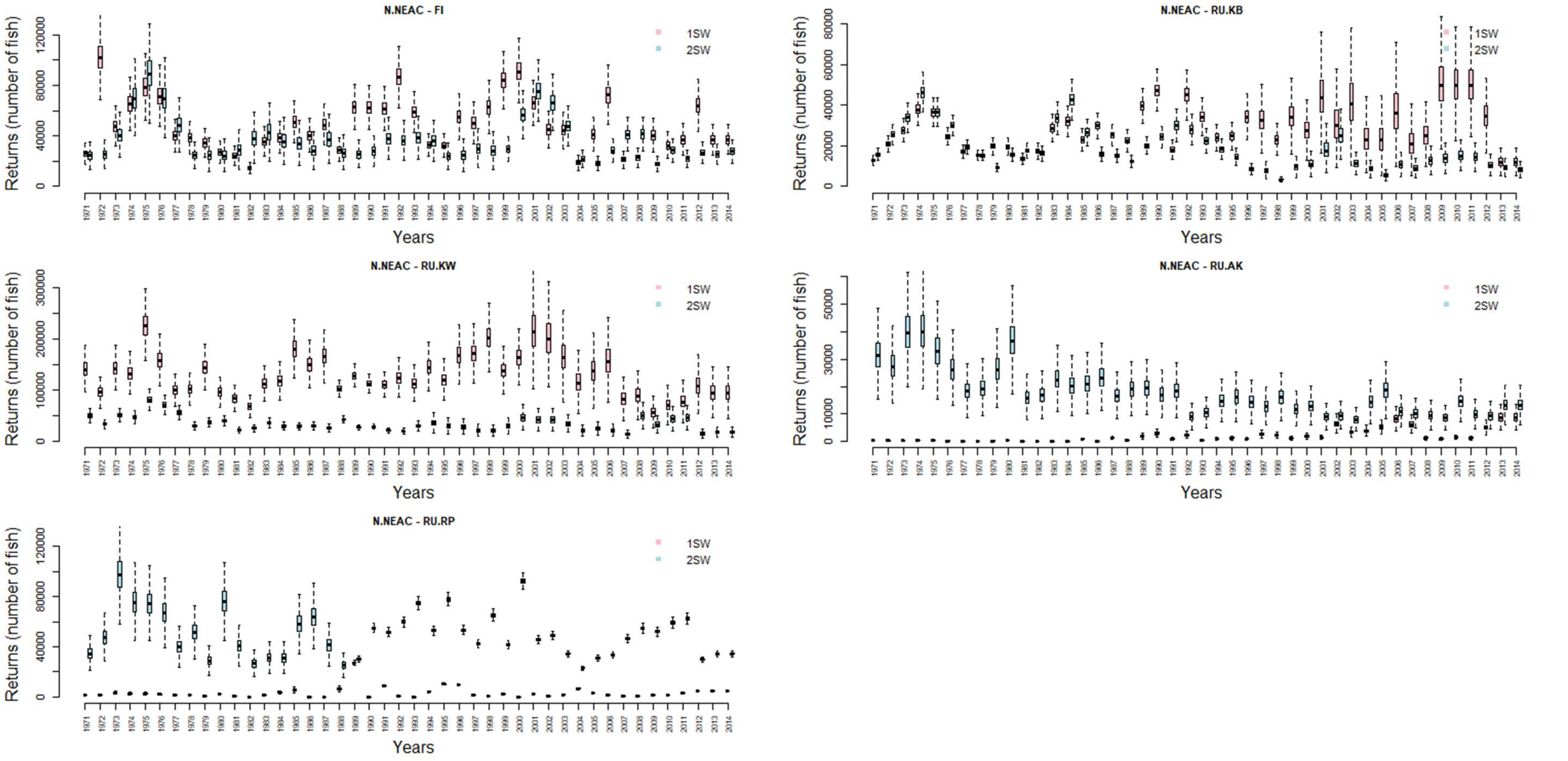

**Figure 4** (Continuing). Probability distributions of the number of fish returning as 1SW (white boxplots) and 2SW (grey boxplots) in each SU of Northern Europe (Source: ICES 2015a).

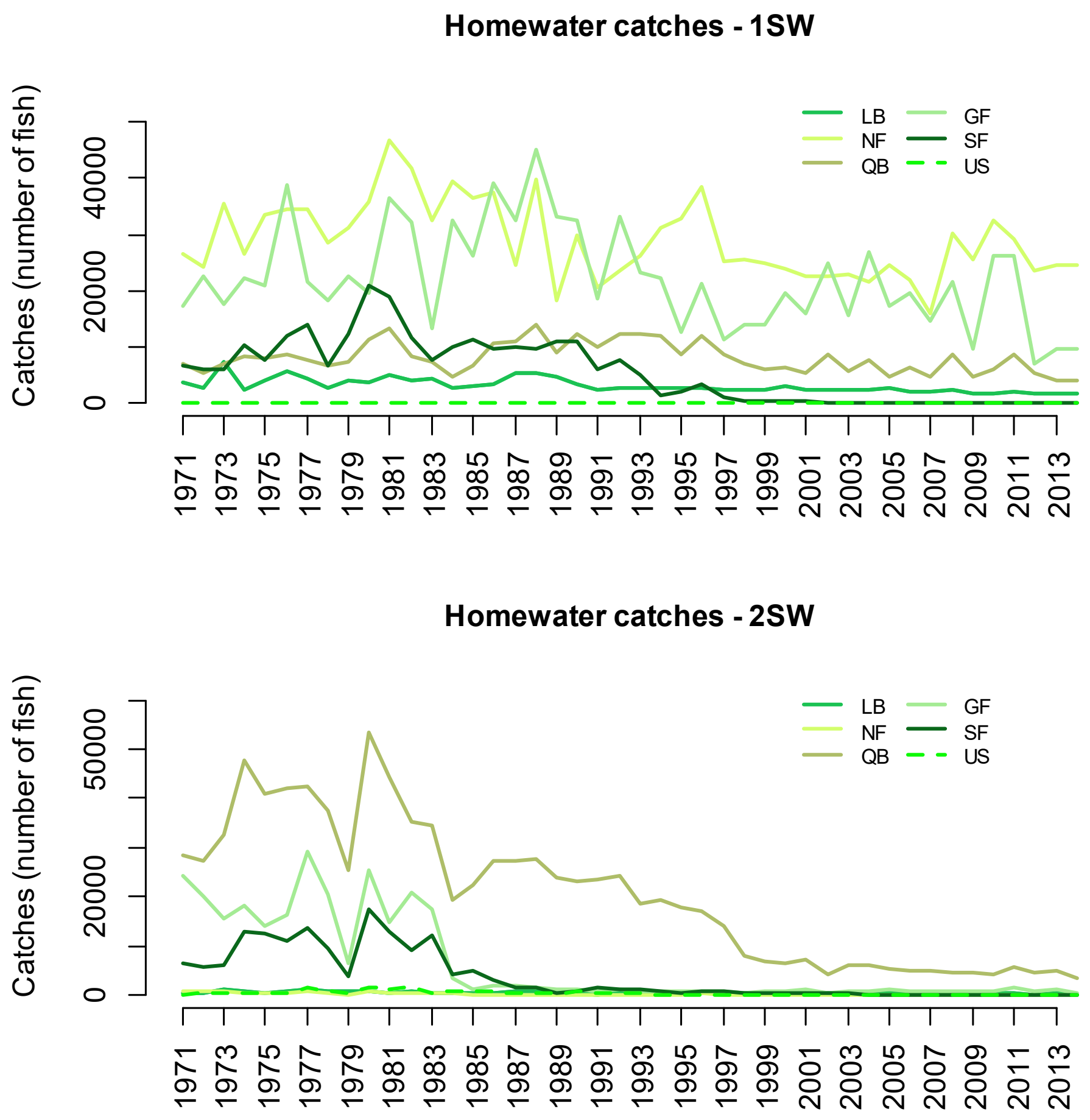

**Figure 5.** Times series of point estimates (median of logNormal probability distributions) of homewater catches for the 6 SU of North America. (a) 1SW fish; (b) 2SW fish (Source: ICES 2015a).

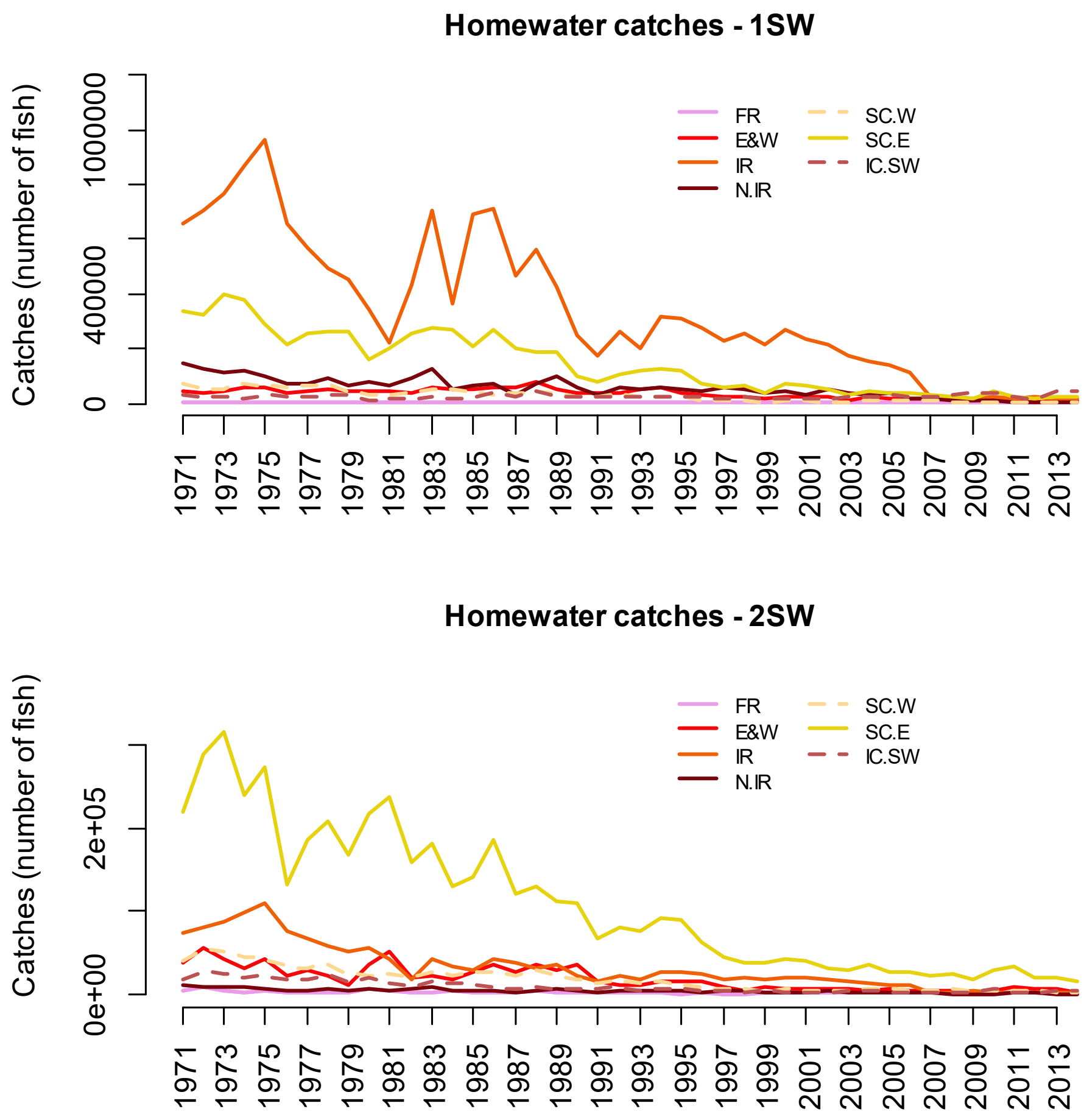

**Figure 5** (continuing). Times series of point estimates (median of logNormal probability distributions) of homewater catches for the 7 SU of Southern Europe. (a) 1SW fish; (b) 2SW fish (Source: ICES 2015a).

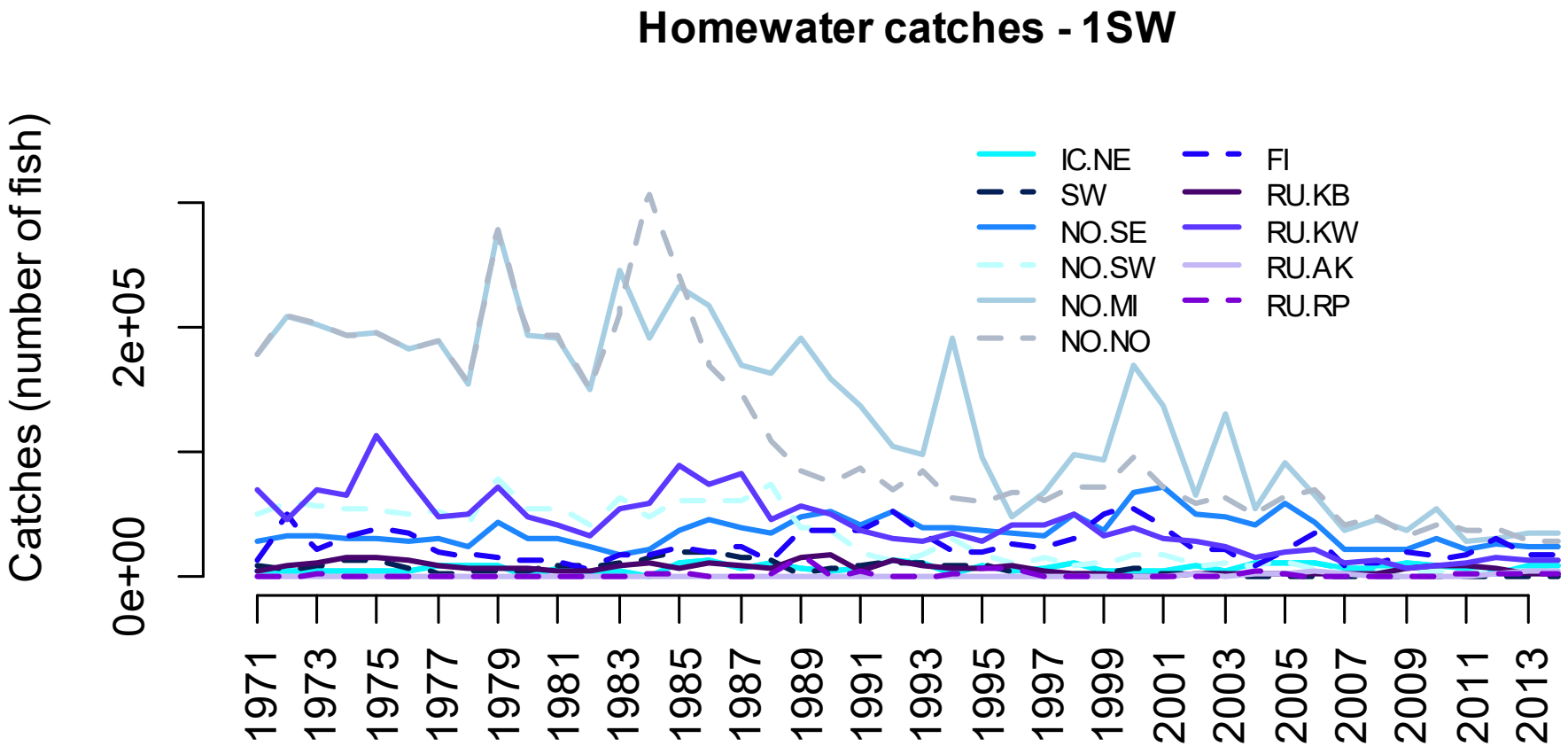

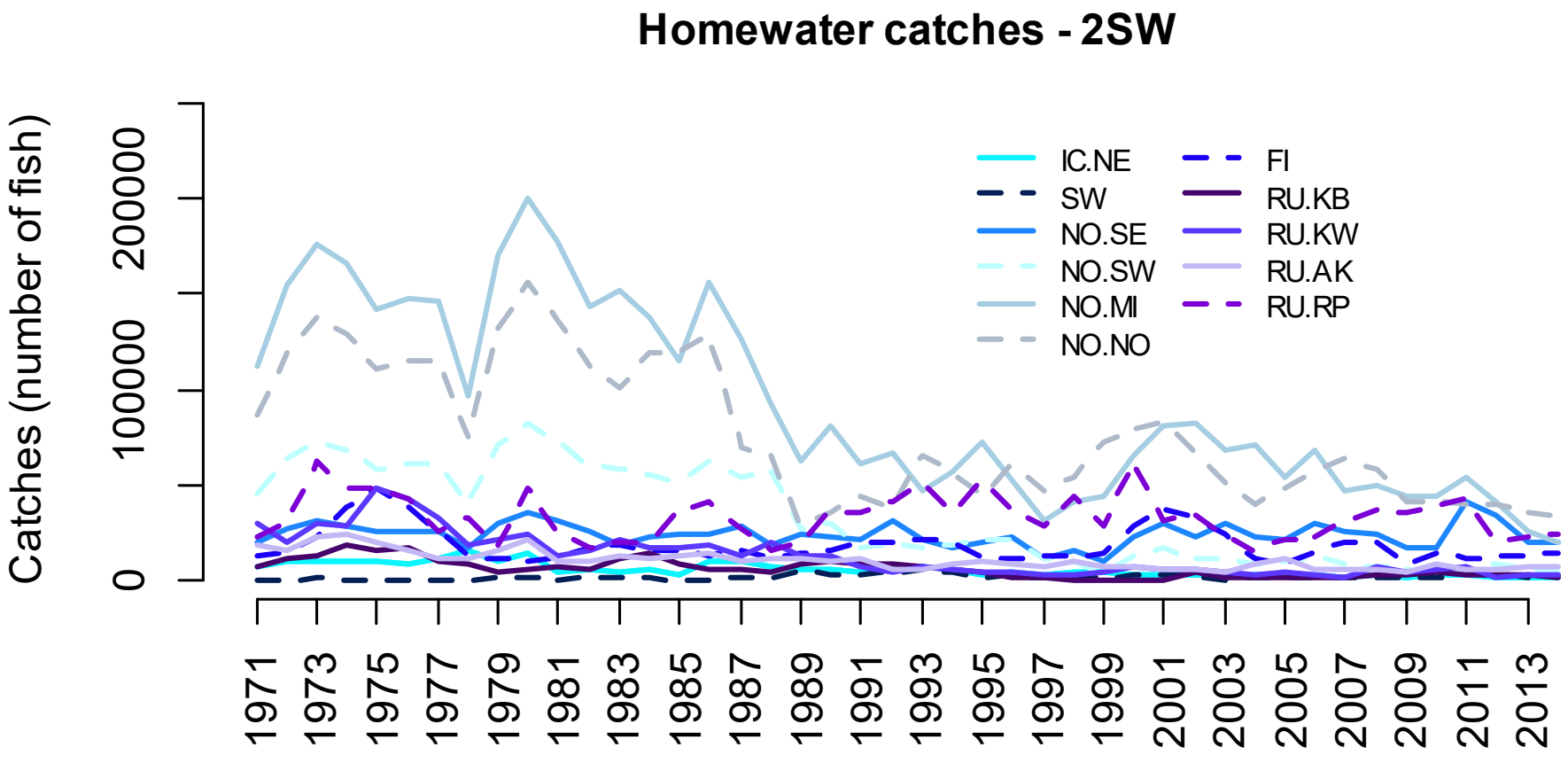

**Figure 5** (continuing). Times series of point estimates (median of logNormal probability distributions) of homewater catches for the 11 SU of Northern Europe. (a) 1SW fish; (b) 2SW fish (Source: ICES 2015a). See text for the hypotheses used to complete the time series for the period 1971-1982.

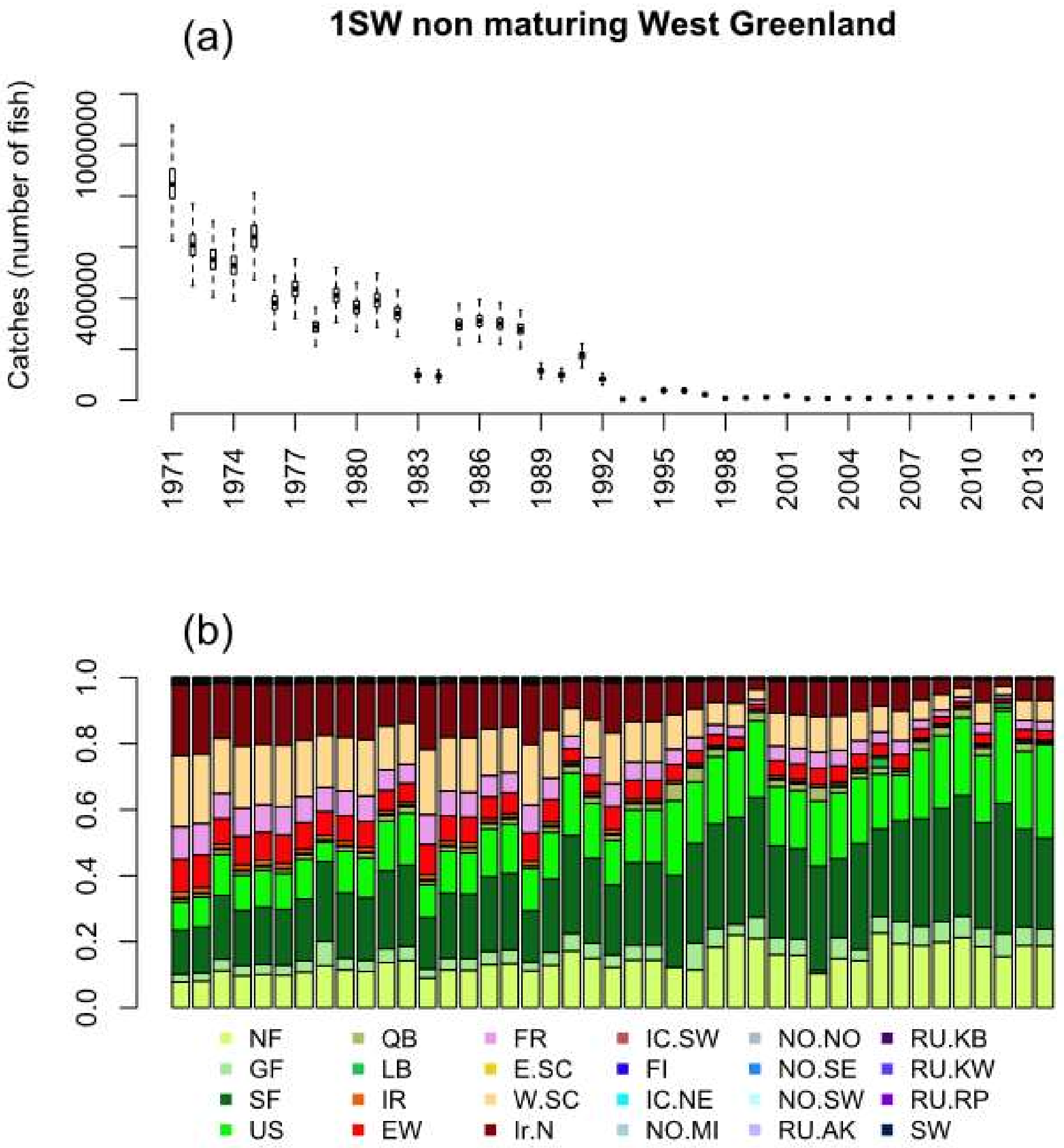

**Figure 6**. (a) Time series of total catches of the 1SW non-maturing stage in the West Greenland fishery (Source: ICES 2015b); (b) proportions of the catches attributed to South European North European and North American stock units (see text for details).

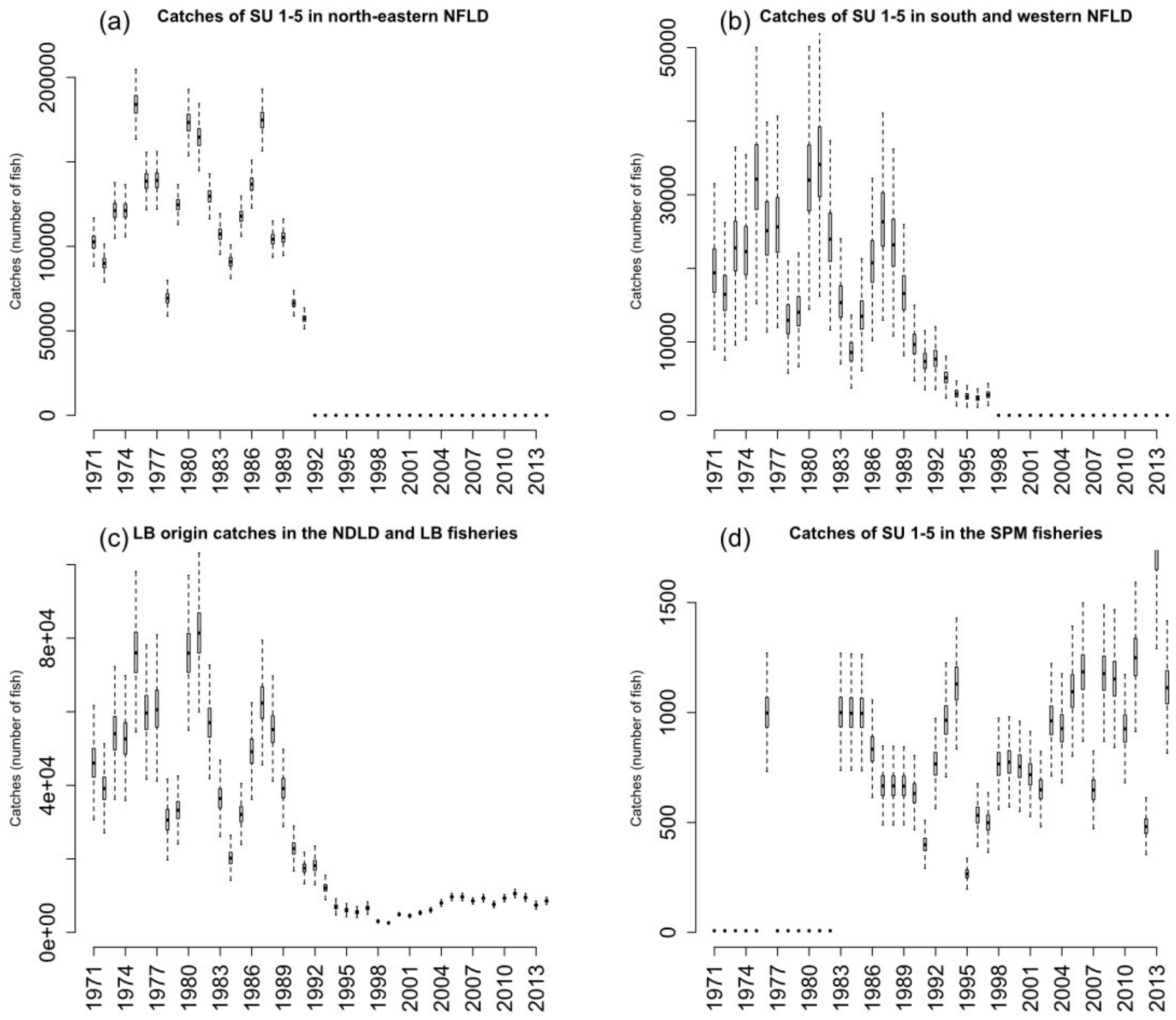

**Figure 7**. Time series of point estimates (median of logNormal distributions) of catches for the sequential fisheries at sea occurring on mixed stocks of North American 1SW maturing fish (Source: ICES 2015a). (a) catches of SU 1-5 (excluding Labrador) in north-eastern Newfoundland (Salmon Fishing Areas 3 to 7); (b) catches of SU 1-5 (excluding Labrador) in south and western Newfoundland (Salmon Fishing Areas 8 to 14A; (c) Labrador (SU 6) origin catches in the Newfoundland and Labrador fisheries (Salmon Fishing Areas 1 to 7); (d) catches of SU 1-5 (excluding Labrador) in the Saint Pierre and Miquelon fisheries.

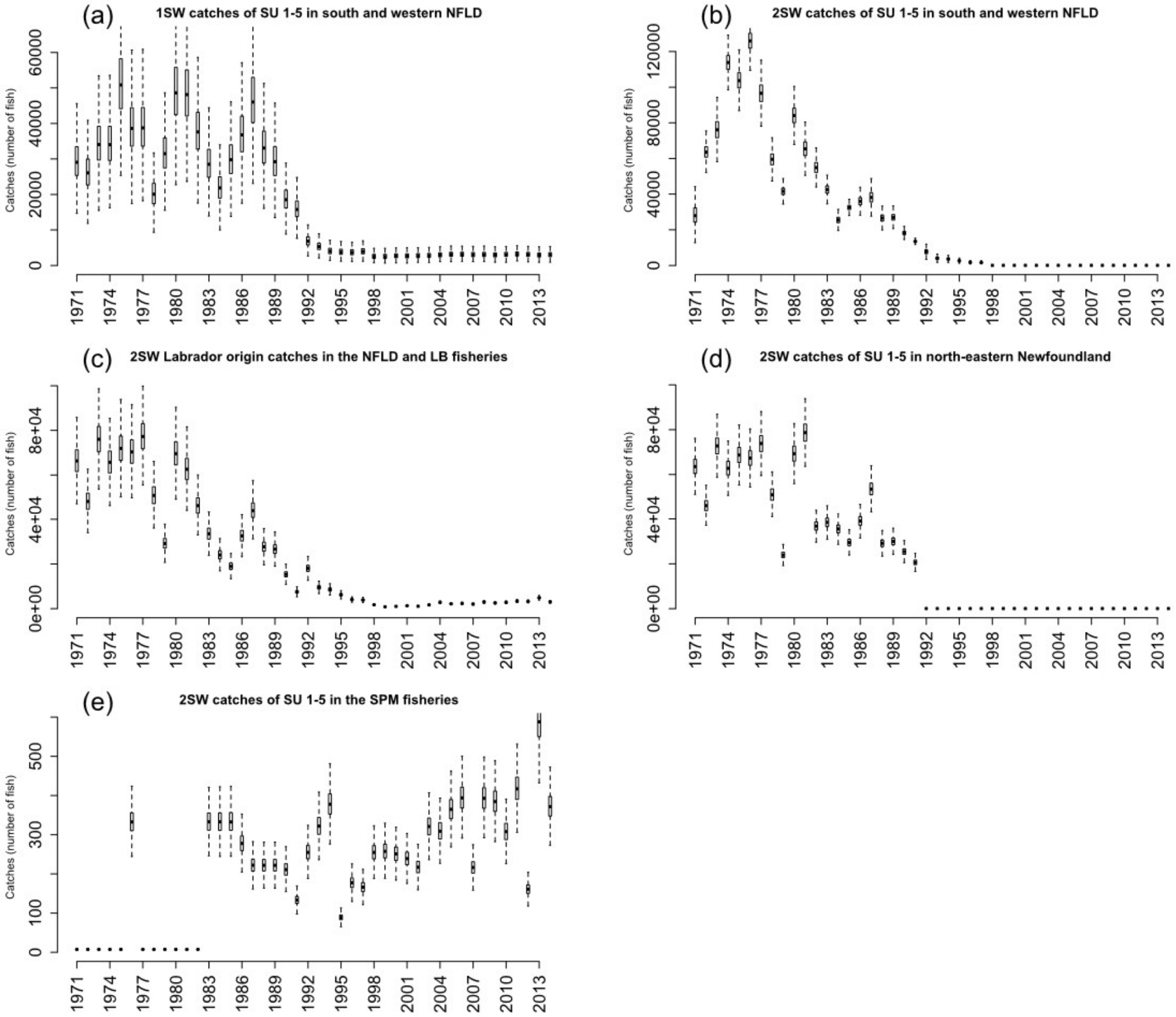

**Figure 7** (Continuing). Time series of point estimates (median of logNormal distributions) of catches for the sequential fisheries at sea occurring on mixed stock fisheries, on North American 1SW non-maturing fish (Source: ICES 2015a). (a) 1SW catches of SU 1-5 (excluding Labrador) in south and western Newfoundland (Salmon Fishing Areas 8 to 14A; (b) 2SW catches of SU 1-5 (excluding Labrador) in south and western Newfoundland (Salmon Fishing Areas 8 to 14A; (c) 2SW Labrador (SU 6) origin catches in the Newfoundland and Labrador fisheries (Salmon Fishing Areas 1 to 7); (d) 2SW catches of SU 1-5 (excluding Labrador) in north-eastern Newfoundland (Salmon Fishing Areas 3 to 7); and (e) 2SW catches of SU 1-5 (excluding Labrador) in the Saint Pierre and Miquelon fisheries.

**Table 6**. Proportions to allocate the total catches among different SU from Southern and Northern Europe in the Faroes fishery. Proportions sum to 1 for each fishery and are considered constant over time (Source: ICES 2015a). Fish originated from from North America are not harvested in the Faroes fishery.

| | S.NEAC | | | | | | | N.NEAC | | | | | | | | | | |
|---|---|---|---|---|---|---|---|---|---|---|---|---|---|---|---|---|---|---|
| | **FR** | **E&W** | **IR** | **N.IR** | **SC.W** | **SC.E** | **IC.SW** | **IC.NE** | **SW** | **NO.SE** | **NO.SW** | **NO.MI** | **NO.NO** | **FI** | **RU.KB** | **RU.KW** | **RU.AK** | **RU.RP** |
| 1SW maturing | 0.021 | 0.082 | 0.341 | 0.070 | 0.107 | 0.249 | 0.014 | 0.005 | 0.001 | 0.015 | 0.003 | 0.026 | 0.018 | 0.010 | 0.008 | 0.027 | 0.002 | 0.001 |
| 1SW non maturing | 0.007 | 0.052 | 0.028 | 0.006 | 0.059 | 0.138 | 0.005 | 0.006 | 0.01 | 0.1 | 0.032 | 0.186 | 0.136 | 0.06 | 0.023 | 0.056 | 0.013 | 0.085 |
| 2SW | 0.007 | 0.052 | 0.028 | 0.006 | 0.059 | 0.138 | 0.005 | 0.006 | 0.01 | 0.1 | 0.032 | 0.186 | 0.136 | 0.06 | 0.023 | 0.056 | 0.013 | 0.085 |

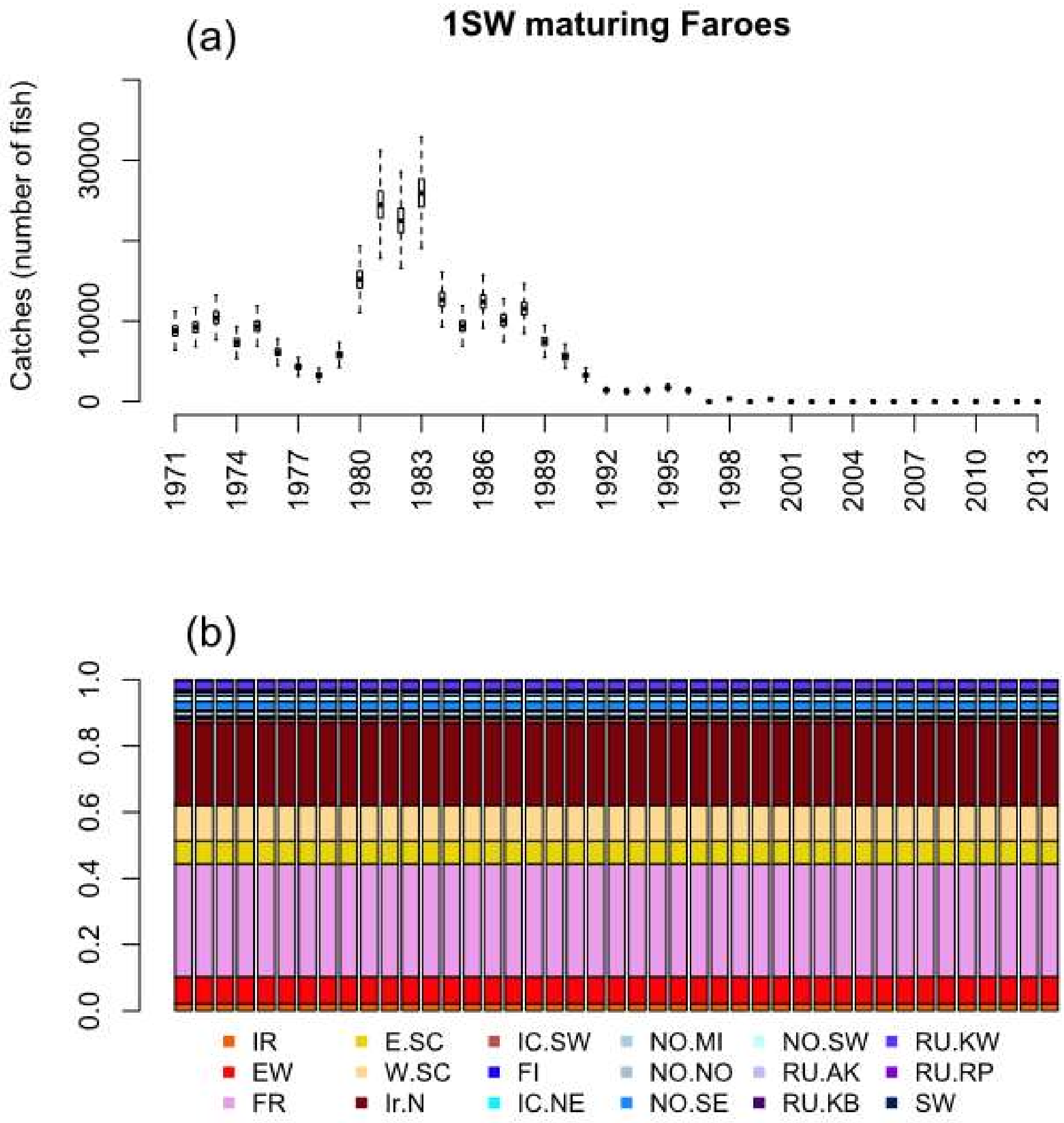

**Figure 8**. (a) Time series of total catches of the 1SW non-maturing stage in the Faroes fishery (Source: ICES 2015b); (b) proportions of the catches attributed to South European and North European stock units (Source: ICES 2015b). (Proportions attributed to SU from NA are 0).

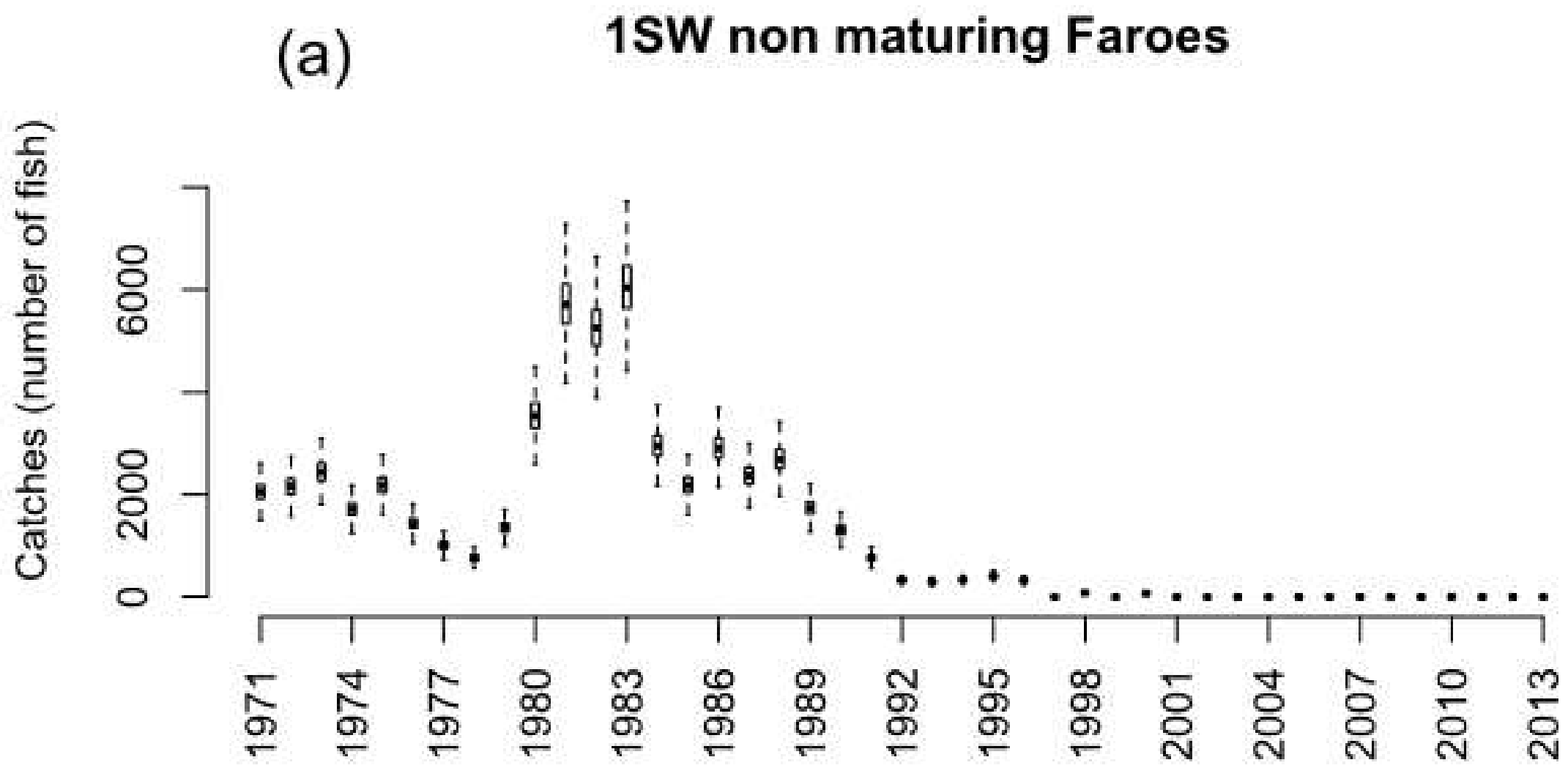

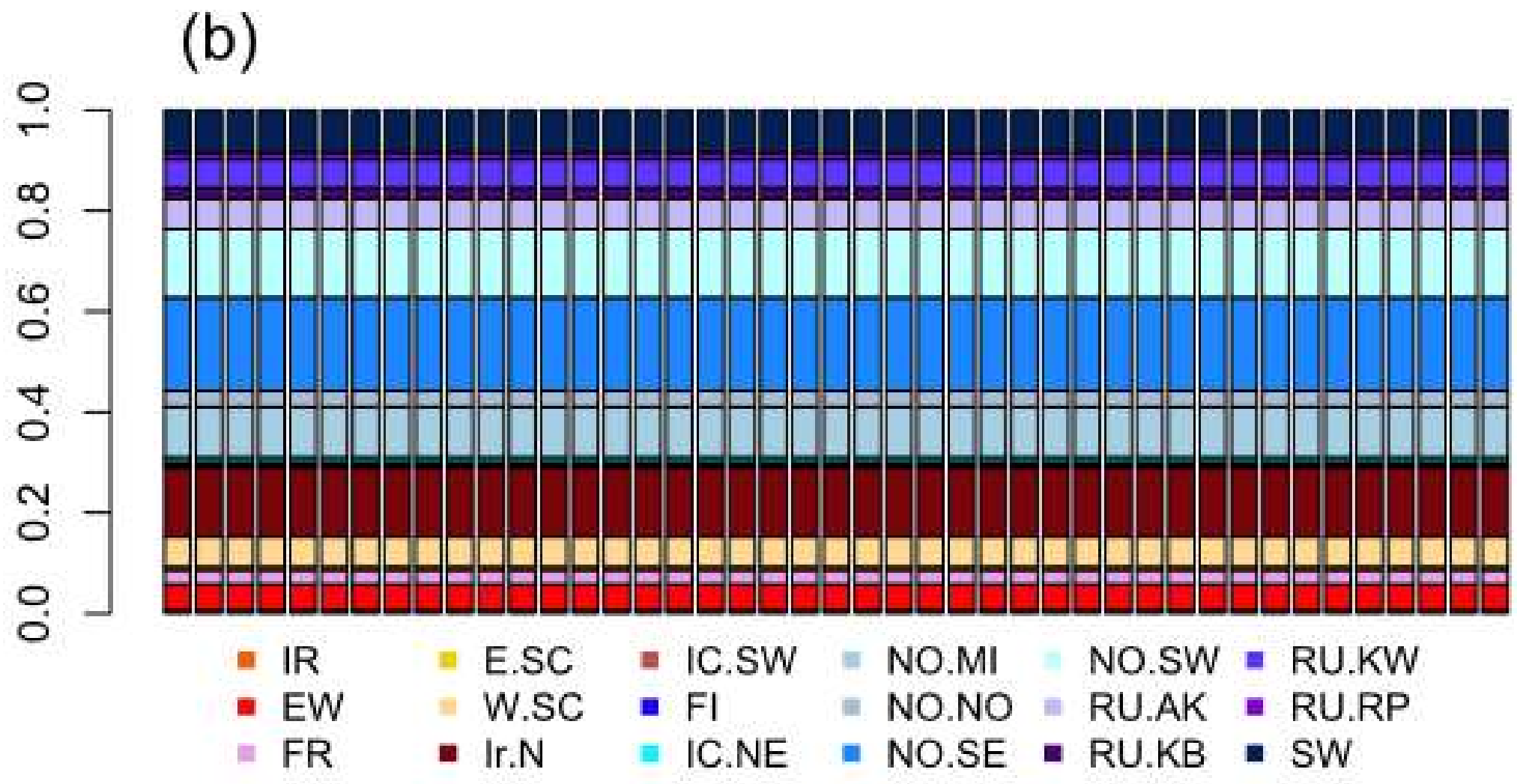

**Figure 8** (continuing). (a) Time series of total catches of the 1SW maturing stage in the Faroes fishery (Source: ICES 2015b); (b) proportions of the catches attributed to South European and North European stock units (Source: ICES 2015b). (Proportion attributed to SU from NA are 0).

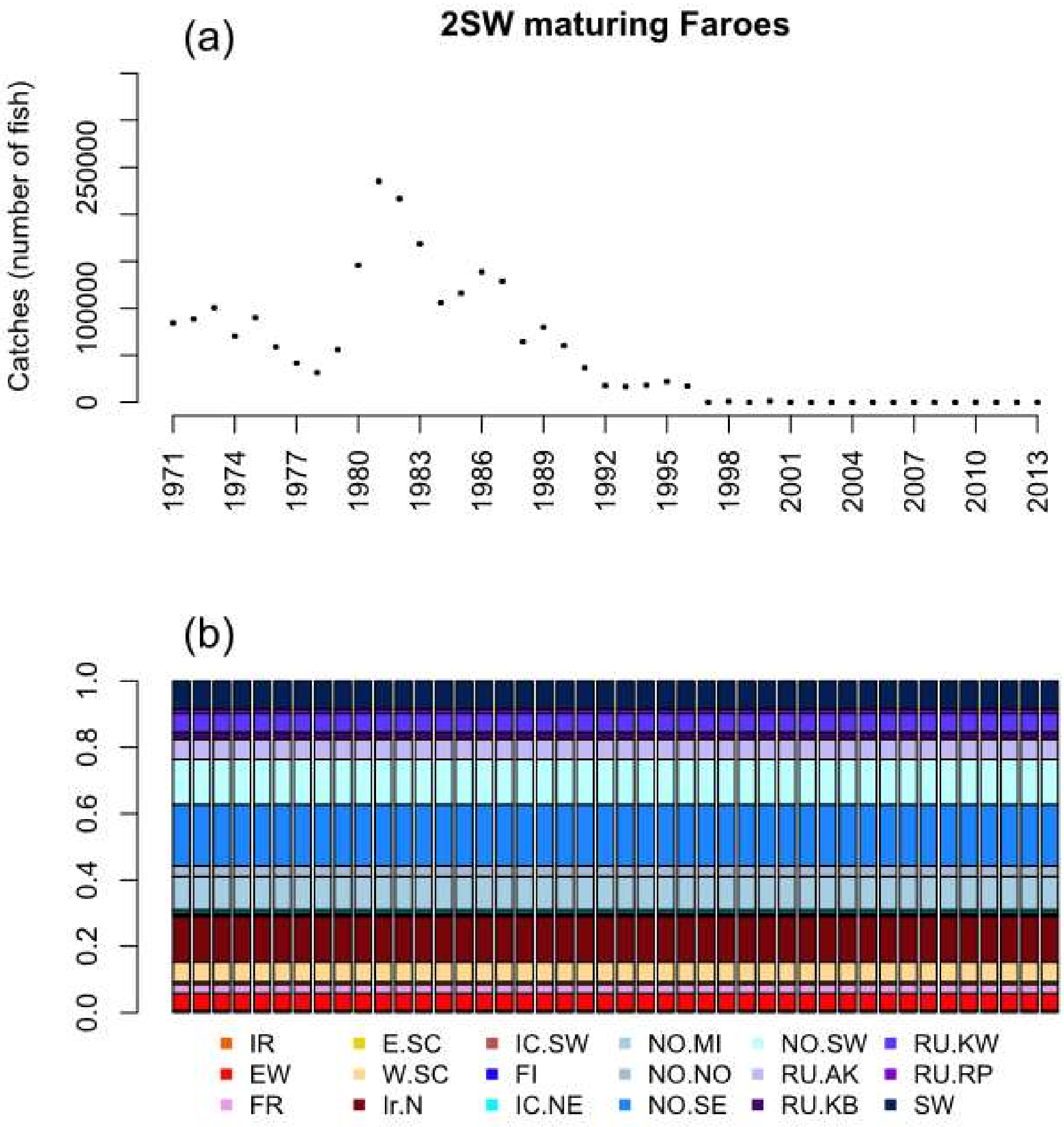

**Figure 8** (continuing). (a) Time series of total catches of the 2SW maturing stage in the Faroes fishery (Source: ICES 2015b); (b) proportions of the catches attributed to South European and North European stock units (Source: ICES 2015b). (Proportion attributed to SU from NA are 0).

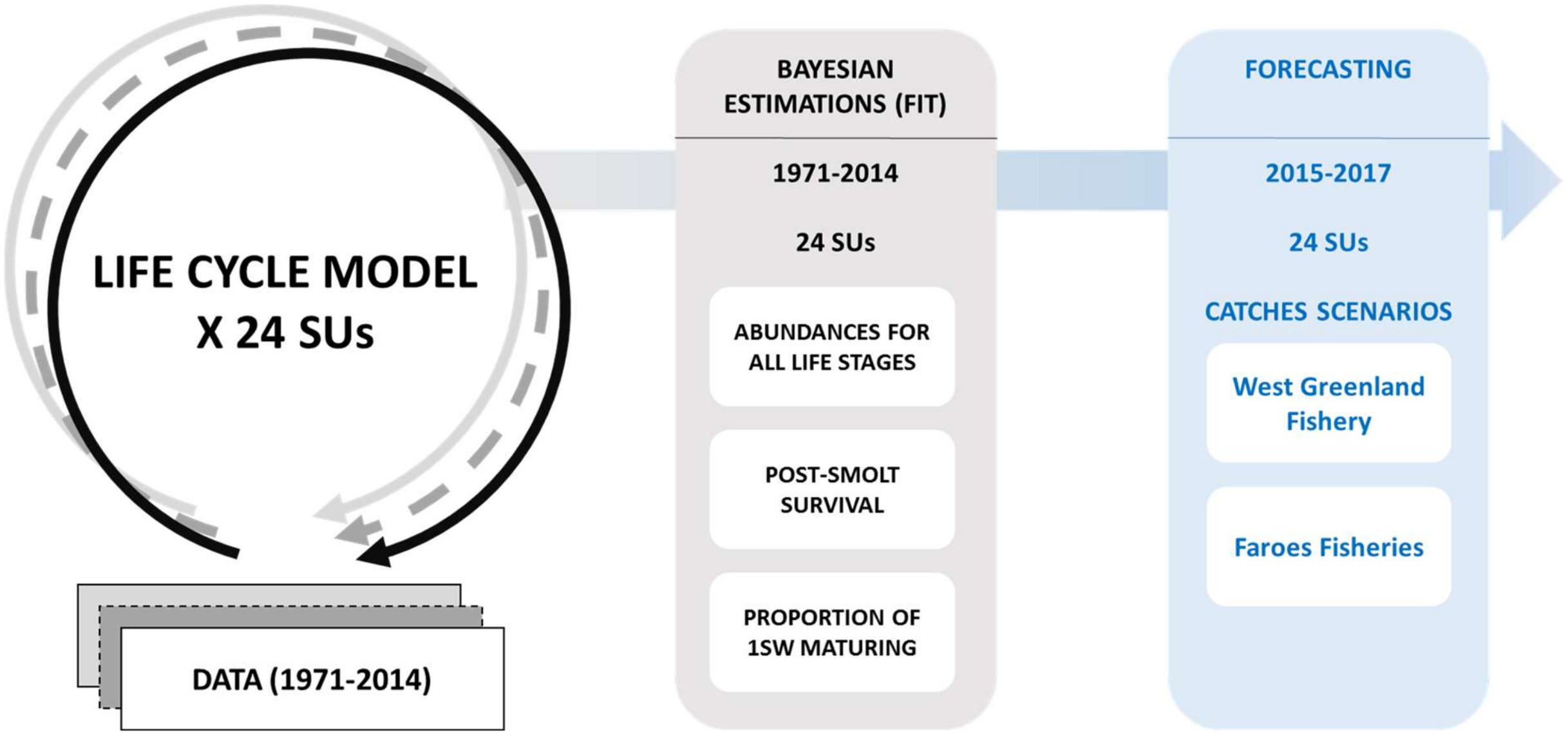

**Figure 9**. Forecasting. The life cycle is first fitted to the time series of data 1971-2013, and then used to forecast abundances for each stock unit (SU) under different scenario of catches in West Greenland and Faroes fisheries. Uncertainty in both the model and the parameters estimates (posterior distribution derived from the fitting phase) are integrated out in the forecasting.

**Table 7**. CL(number of eggs) used for the development of catch options for the stocks units in North America, Southern Europe and Northern Europe.

| Stock Units | CLs | References |
|---|---|---|
| North America | | |
| Labrador | 243660000 | O'Connel et al. 1997 |
| Newfoundland | 267780000 | Reddin et al. 2009 |
| Quebec | 50380000 | Atlantic salmon management plan 2016, Ministère des Forêts, de la Faune et des Parcs (2016). |
| Gulf | 248680000 | Cameron et al. 2009, Breau et al. 2009, Chaput et al. 2010, Cairns et al. 2015 |
| Scotia Fundy | 224140000 | Gibson et al. 2014, Bowlby et al. 2013, Jones et al. 2014 |
| US | 435369000 | Baum, E.T. 1995 |
| Southern Europe | | |
| Iceland (south+west) | 64273104 | ICES, 2015a |
| Scotland | 1609542000 | |
| Northern Ireland | 56281942 | |
| Ireland | 710711690 | |
| England&Wales | 211419850 | |
| France | 55165500 | |
| Northern Europe | | |
| Iceland (north+east) | 23889096 | ICES, 2015a |
| Sweden | 13997100 | |
| Norway | 444064980 | |
| Finland | 104278220 | |
| Russia | 357856550 | |

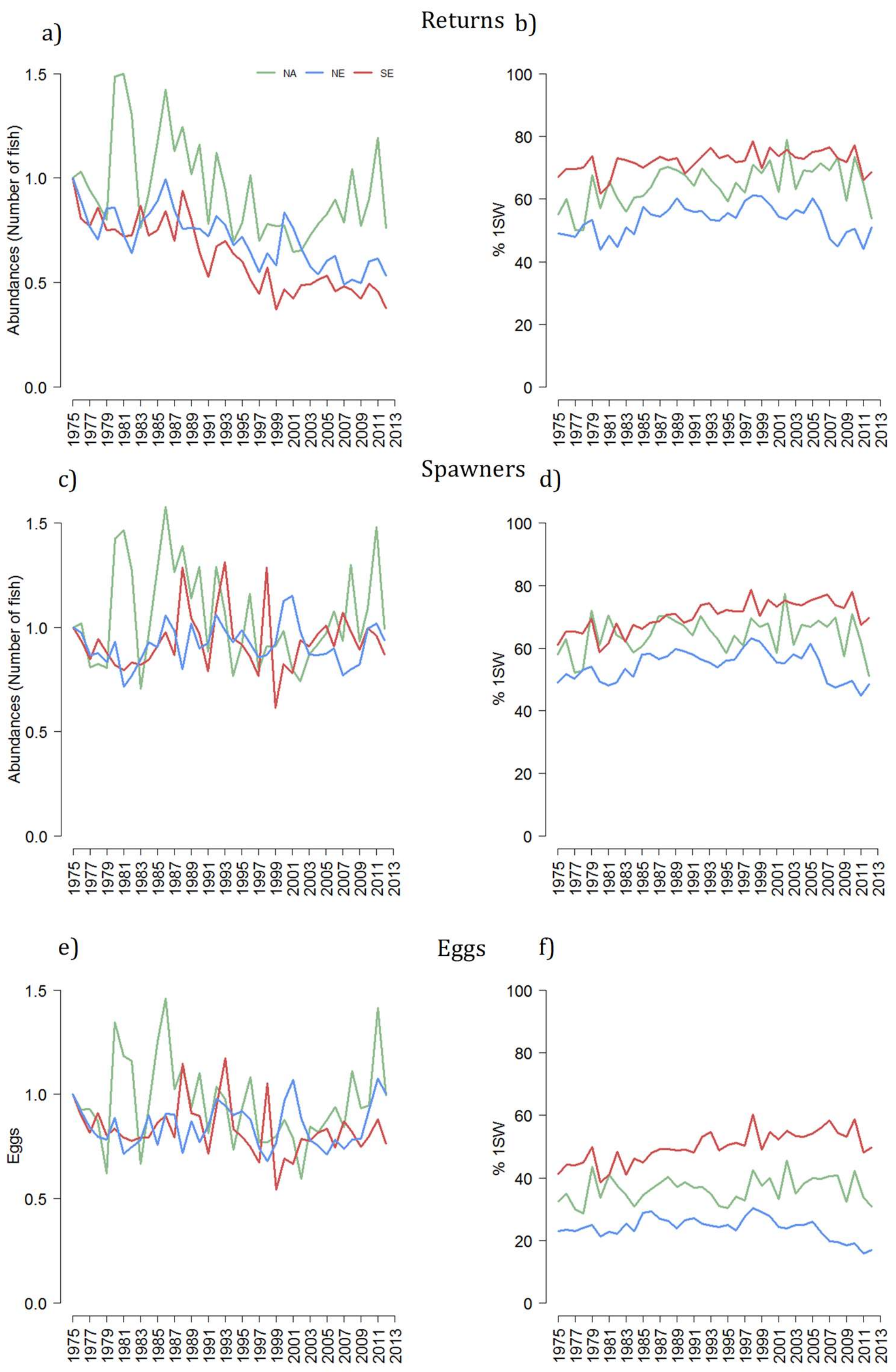

**Figure 10**: Time series of estimated abundances averaged per CSG at four stages in the life cycle. (a) total returns to homewater (1SW + 2SW); (b) proportion of 1SW in returns; (c) total spawners (1SW + 2SW); (d) proportion of 1SW in spawners; (e) total egg deposition by spawners; (f) proportion of eggs spawned by 1SW. Trend lines are medians of marginal posterior distributions. Abundances are standardized to the first year values.

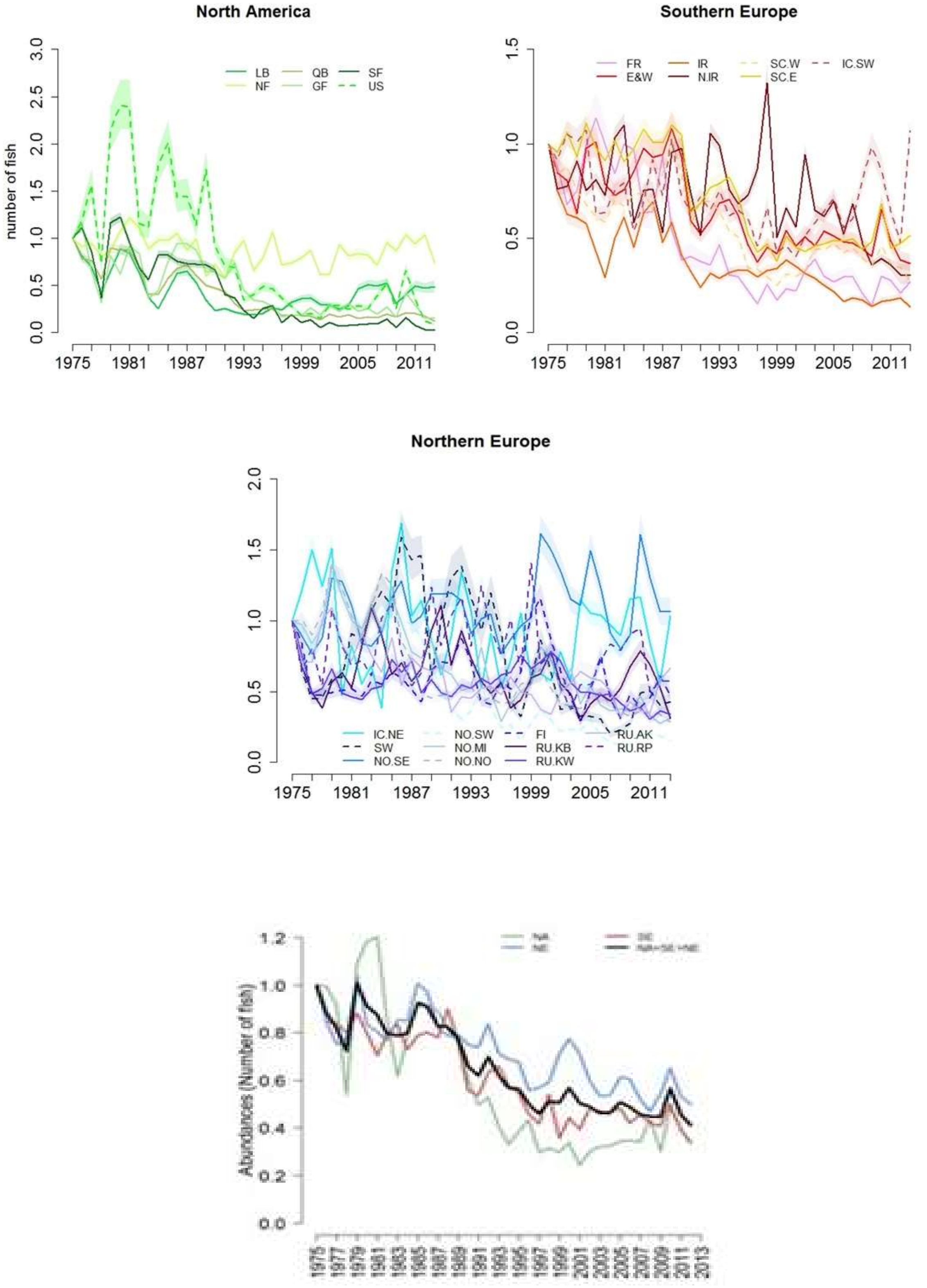

**Figure 11.** Time series of estimated abundances at the PFA stage (maturing + non maturing PFA) for all SU for the three continental stock groups and summed by CSG (bottom panel). Thick lines: median of the marginal posterior distributions. Shaded areas: 50% posterior credibility intervals. PFA are standardized to the first year values.

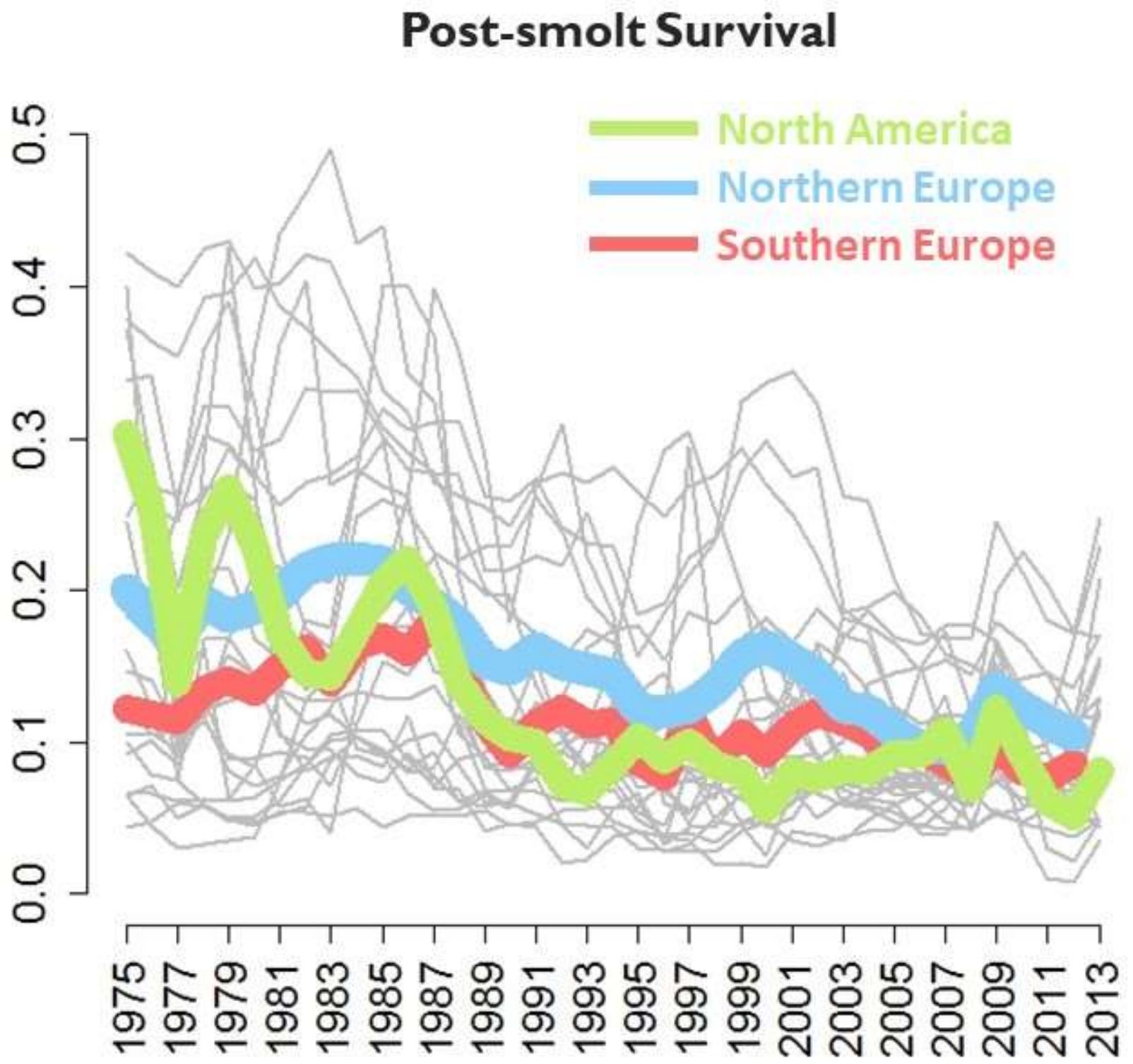

**Figure 12**. Time series of smolt-PFA survival (plotted in the natural scale) for the 24 SU (thin grey lines) and averaged over the three continental stock groups (thick color lines).

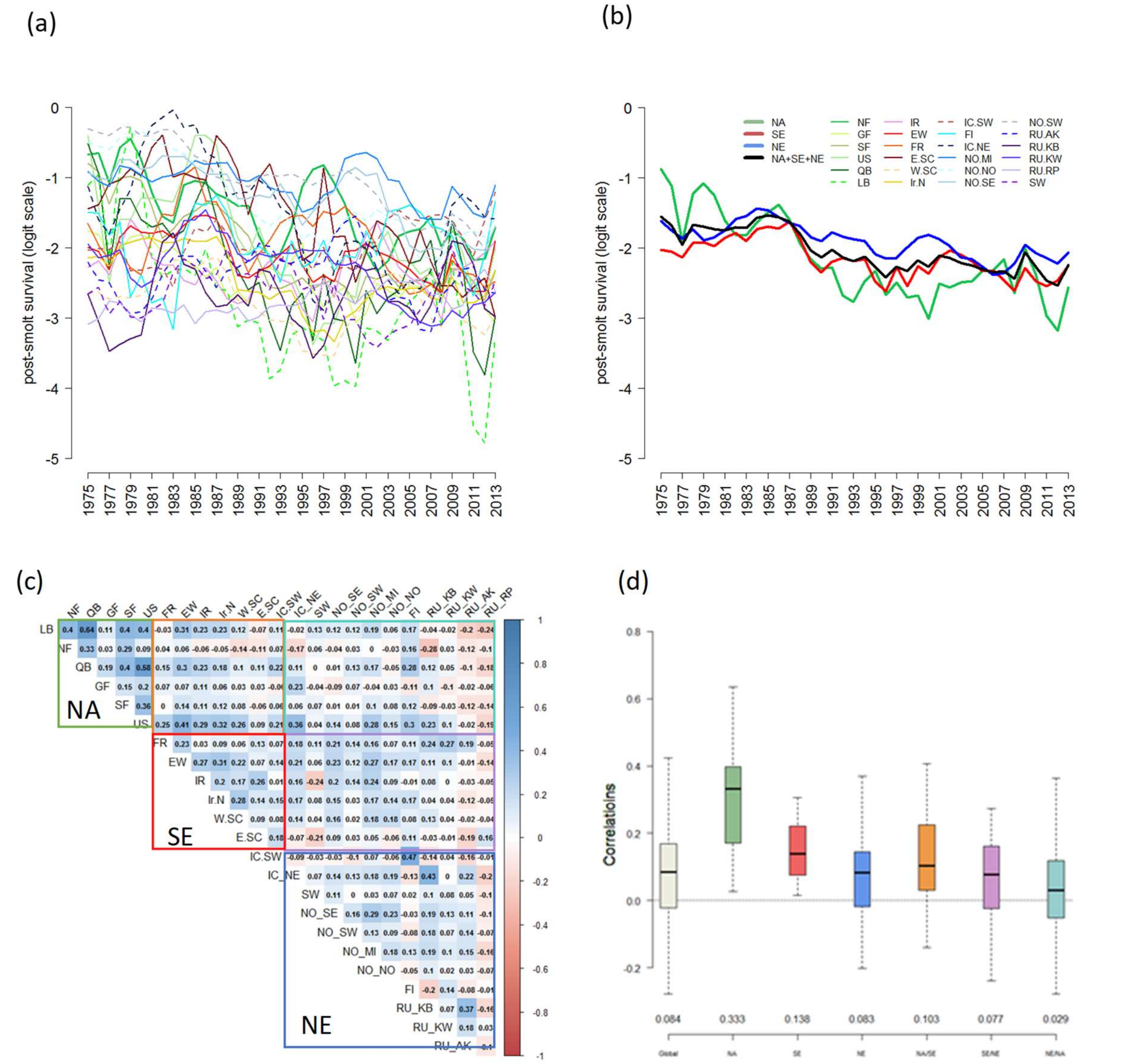

**Figure 13**. Time trends ((a) and (b)) and covariance ((c) and (d)) in the post smolt survival. (a) Time series of post-smolt survivals (logit scale) estimated for the 24 SU (medians of marginal posterior distributions). (b) Average post-smolt survival (logit scale) calculated over all SU in the same CSG (NA: green, SE: red, NE: blue). (c) Pairwise correlations calculated between all SUs. (d) Pairwise correlations averaged over all SUs, over SU within the same CSG (NA, SE, NE) and over pairs of SU that belong to two different CSG (color rectangles). Values on the bottom line indicate the median pairwise correlations.

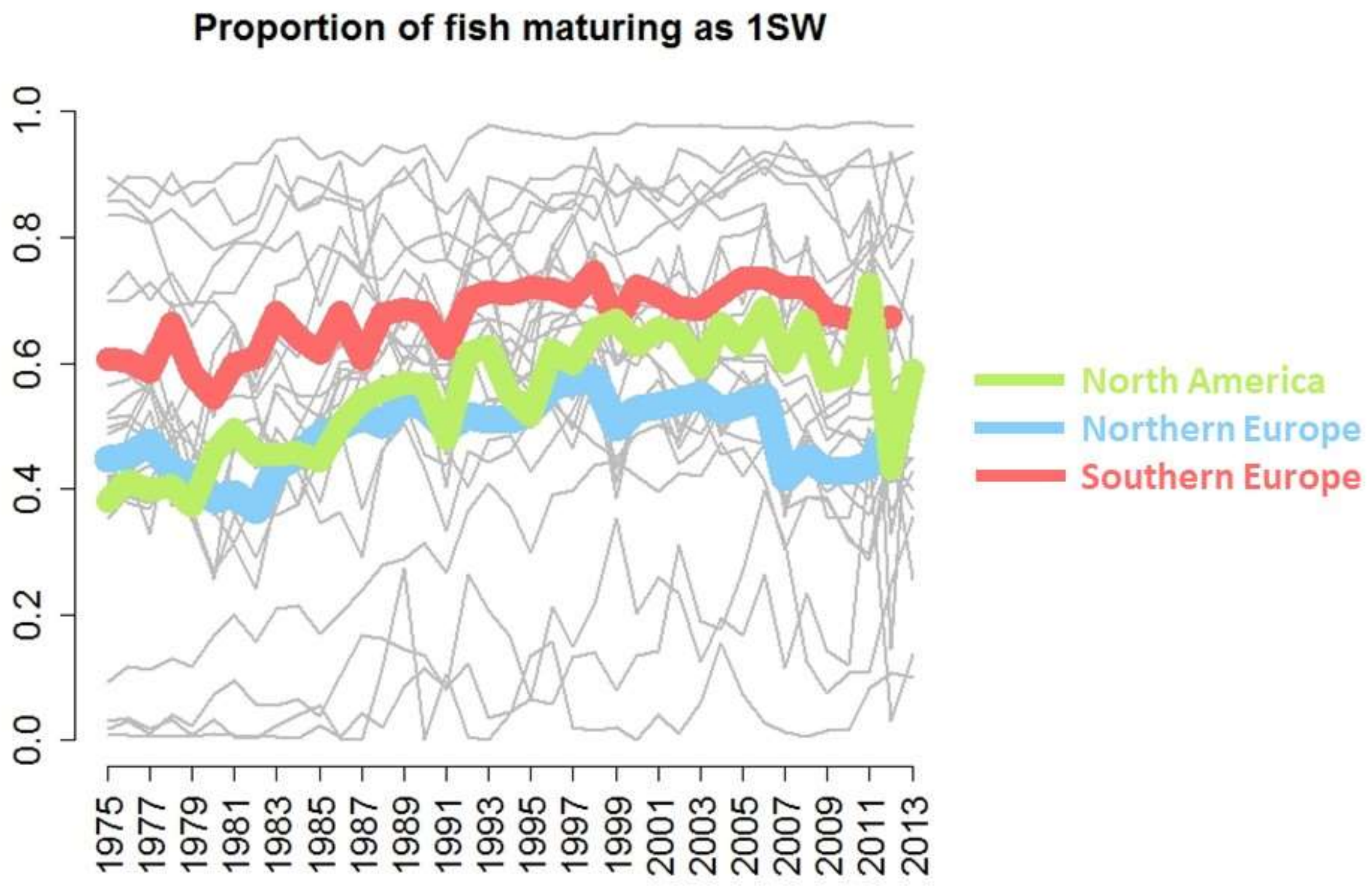

**Figure 14**. Time series of proportion of fish maturing as 1SW (plotted in the natural scale) for the 24 SU (thin grey lines) and averaged over the three continental stock groups (thick color lines). Blue: Northern Europe.

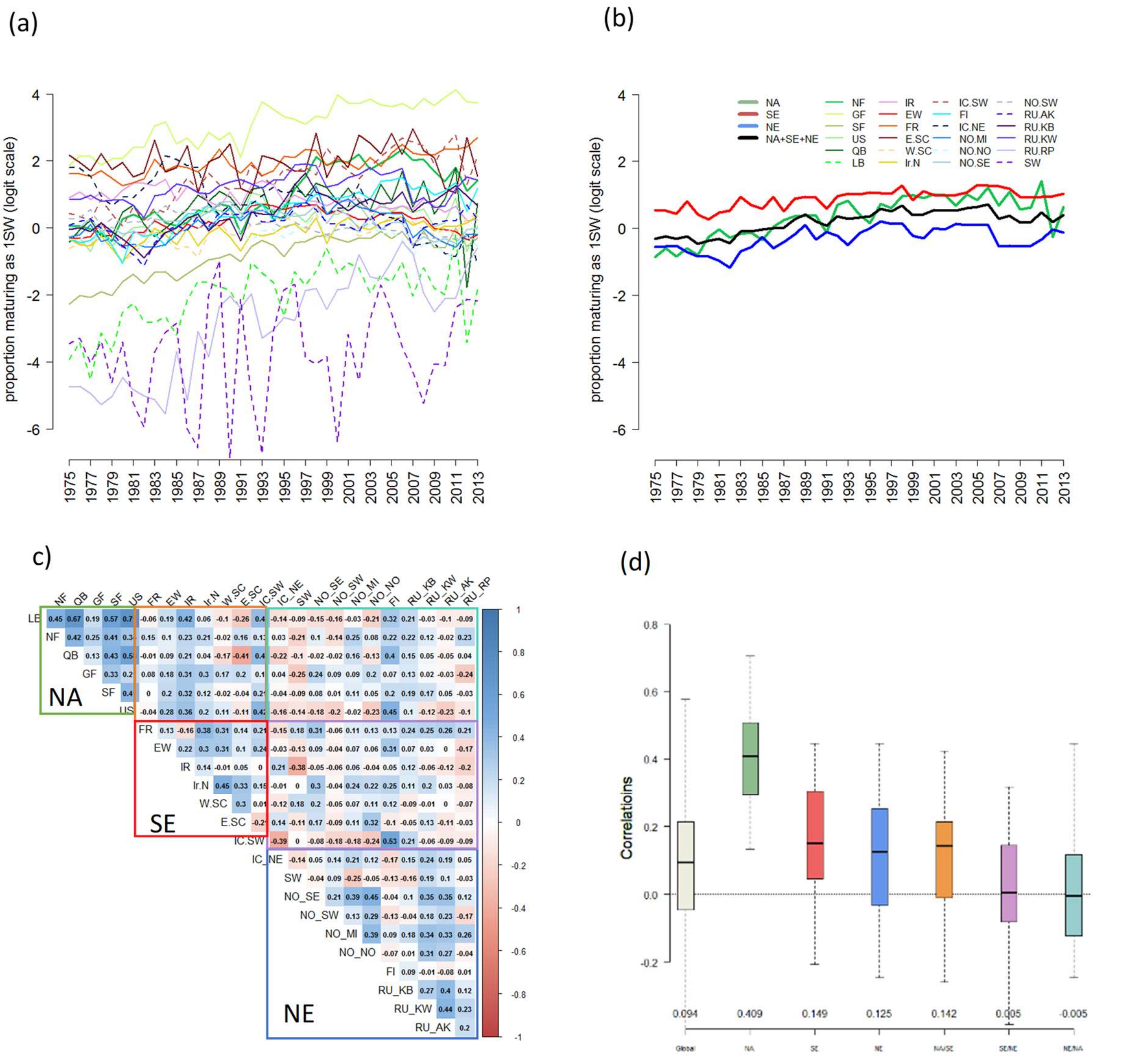

**Figure 15**. Time trends ((a) and (b)) and covariance ((c) and (d)) in the proportion of fish maturing as 1SW. (a) Time series of proportion maturing as 1SW (logit scale) estimated for the 24 SU (medians of marginal posterior distributions). (b) Proportion of fish maturing as 1SW (logit scale) averaged over SU in the same CSG (NA: green, SE: red, NE: blue). (c) Pairwise correlations calculated between all SU. (d) Pairwise correlations averaged between all SU, over SU within the same CSG (NA, SE, NE) and over pairs of SU that belong to two different CSG (colour rectangles). Values on the bottom line indicate the median pairwise correlations.

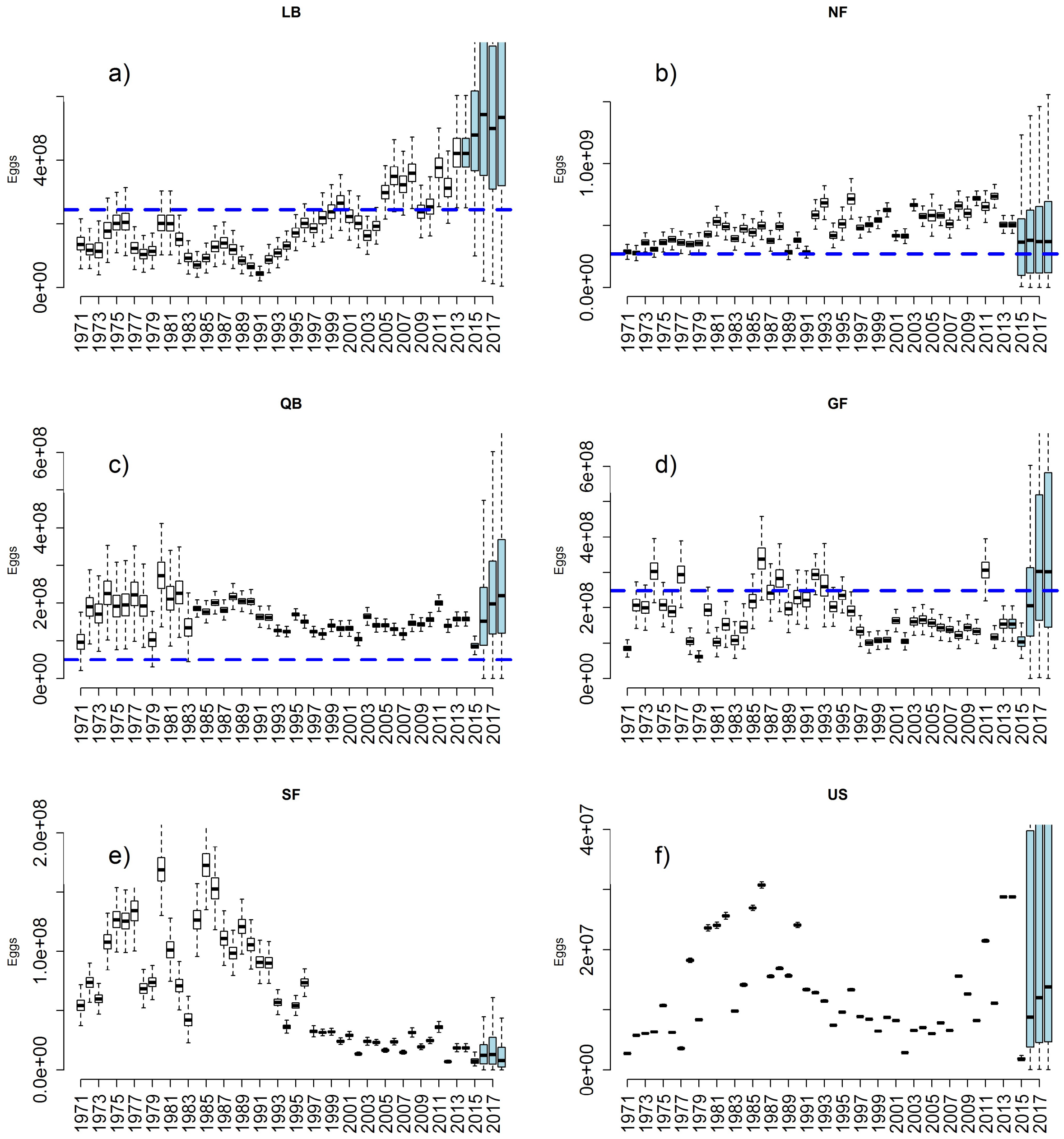

**Figure 16**. Probability distributions of the number of egg potentially spawned for all SU or aggregate of SU. White boxplots: historical time series; Blue boxplots: forecast. (a-f) SU of North America.

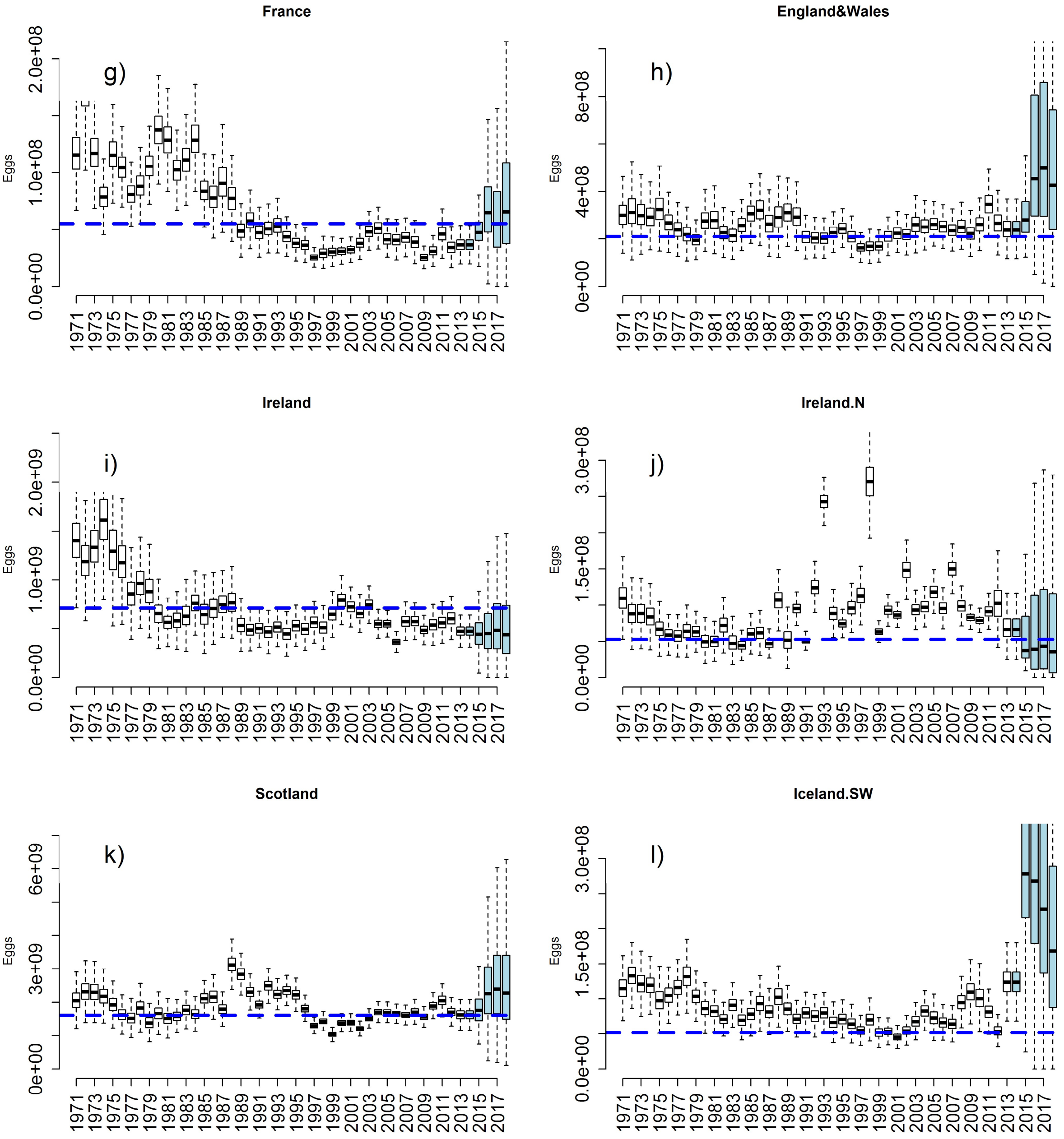

**Figure 16** (continuing). Probability distributions of the number of egg potentially spawned for all SU or aggregate of SU. White boxplots: historical time series; Blue boxplots: forecast. (g-l) SU of Southern Europe.

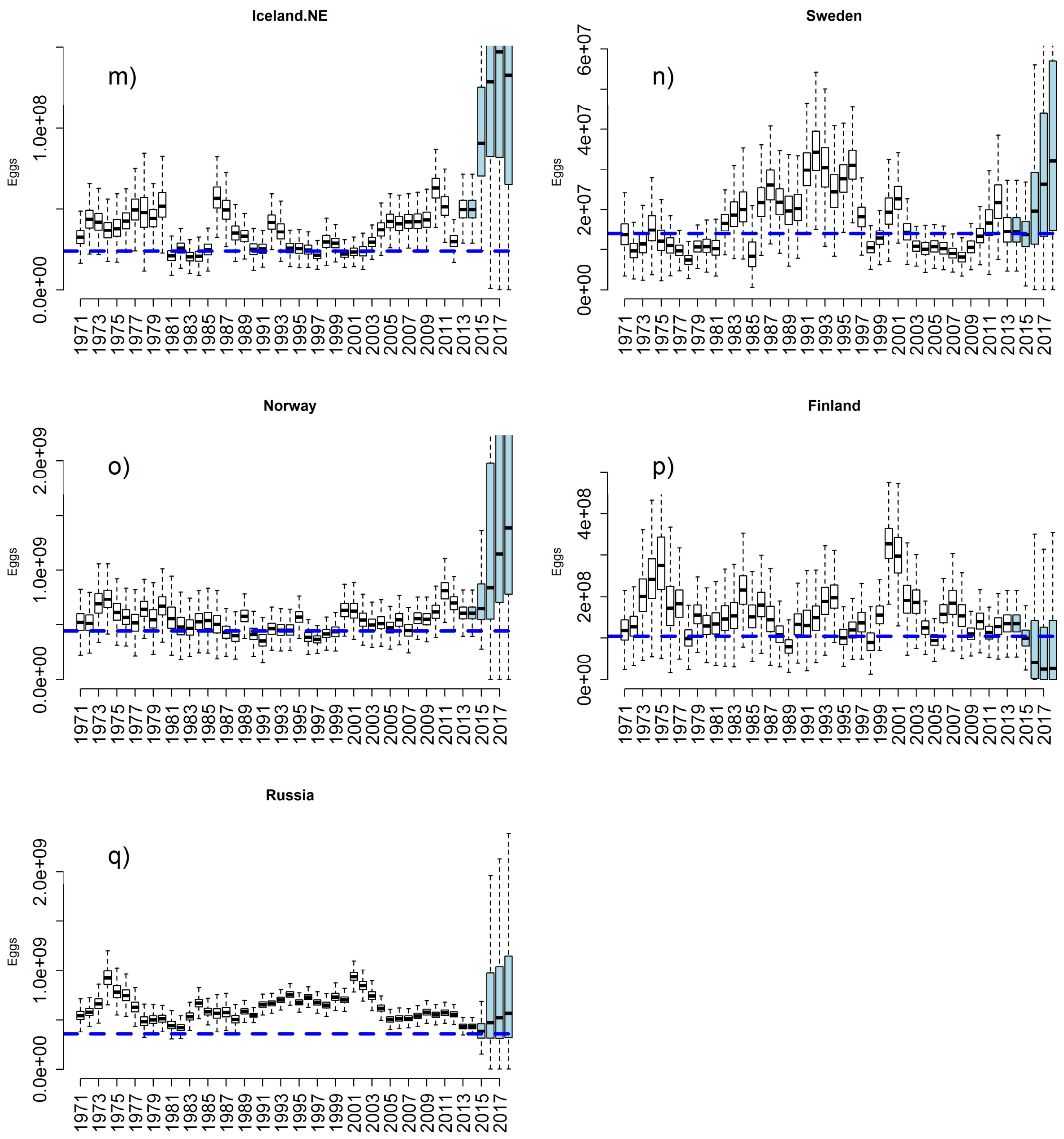

**Figure 16** (continuing). Probability distributions of the number of egg potentially spawned for all SU or aggregate of SU. White boxplots: historical time series; Blue boxplots: forecast. (l-p) Stock units of Northern Europe.

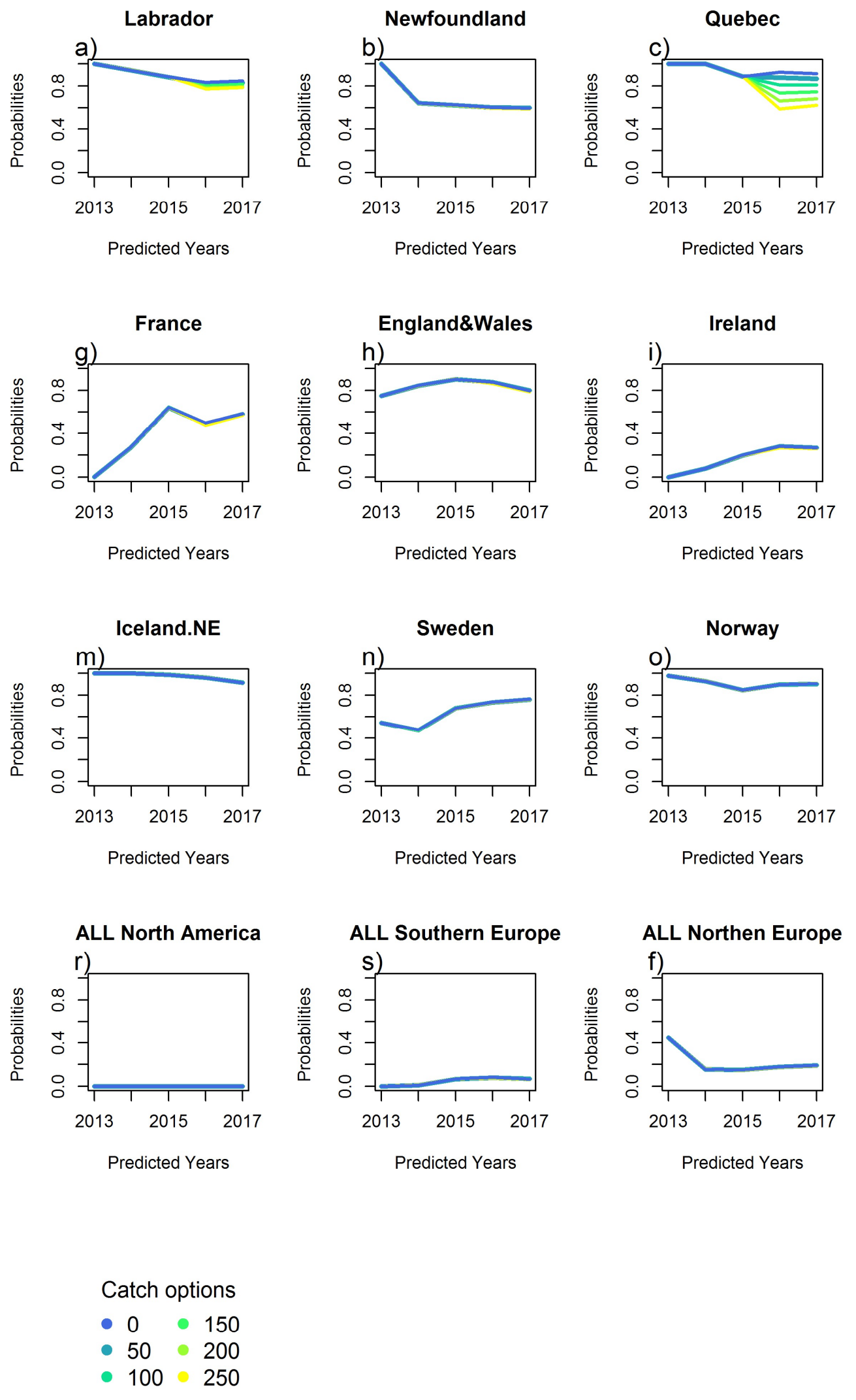

**Figure 17**. Probability to reach Conservation Limits obtained under different catches options at West Greenland. Catches options: 0, 50, 100, 150, 200 and 250 tons (5 years projections). (a-f) North America; (g-l) Southern Europe; (l-p) Northern Europe. Pannels (r-s-t) give probabilities to simultaneously achieving the management objectives for all SU of North America (r), Southern Europe (s) and Northern Europe (t).

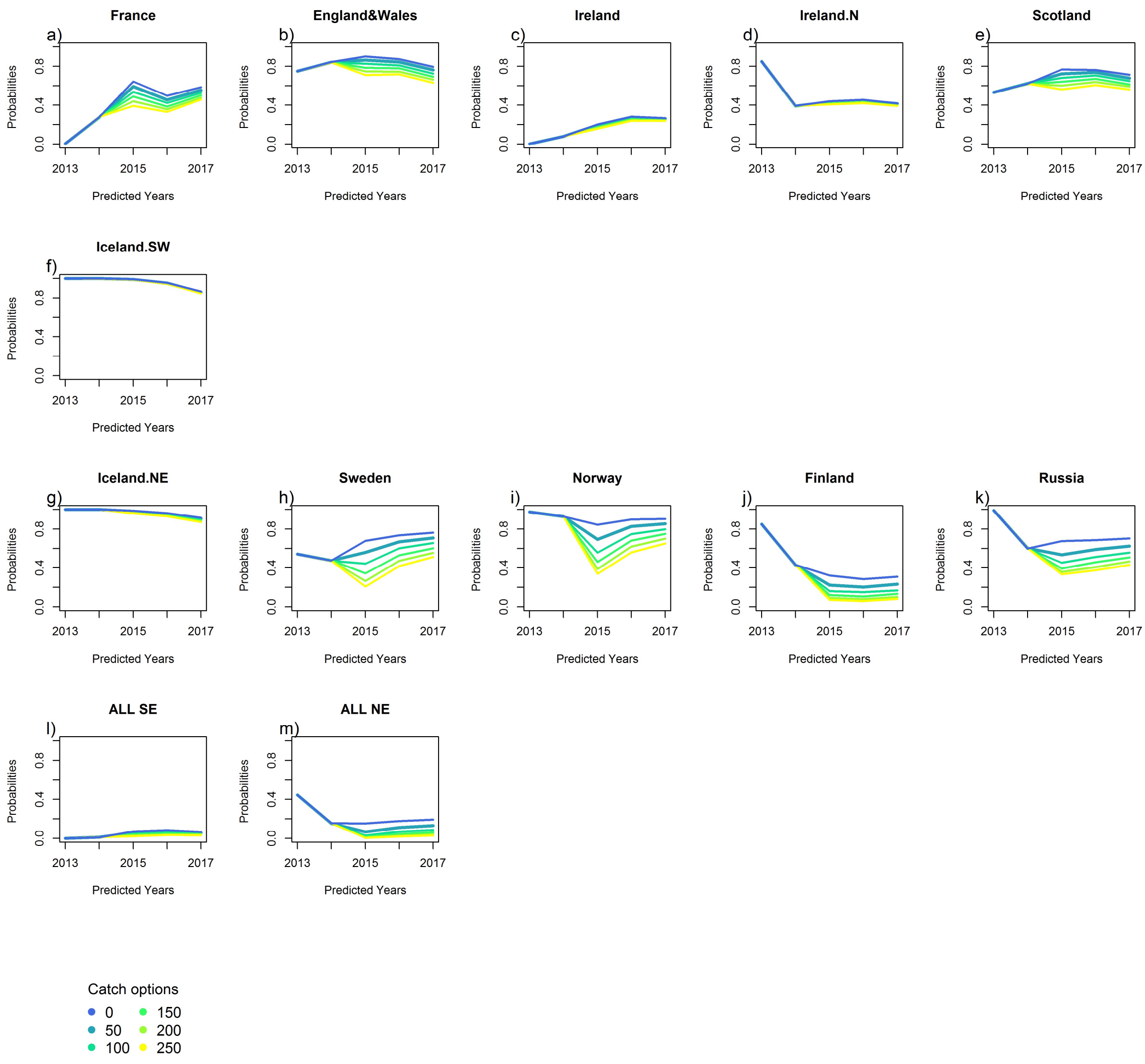

**Figure 18**. Probability to reach Conservation Limits under different catches options at Faroes. Catches options: 0, 50, 100, 150, 200 and 250 tons (5 years projections). (a-f) Southern Europe; (g-k) Northern Europe. Pannels (l-m) give probabilities to simultaneously achieving the management objectives for all SU of Southern Europe (l) and Northern Europe (m). Stock Units of North America are not impacted by Faroes fisheries and are not represented.

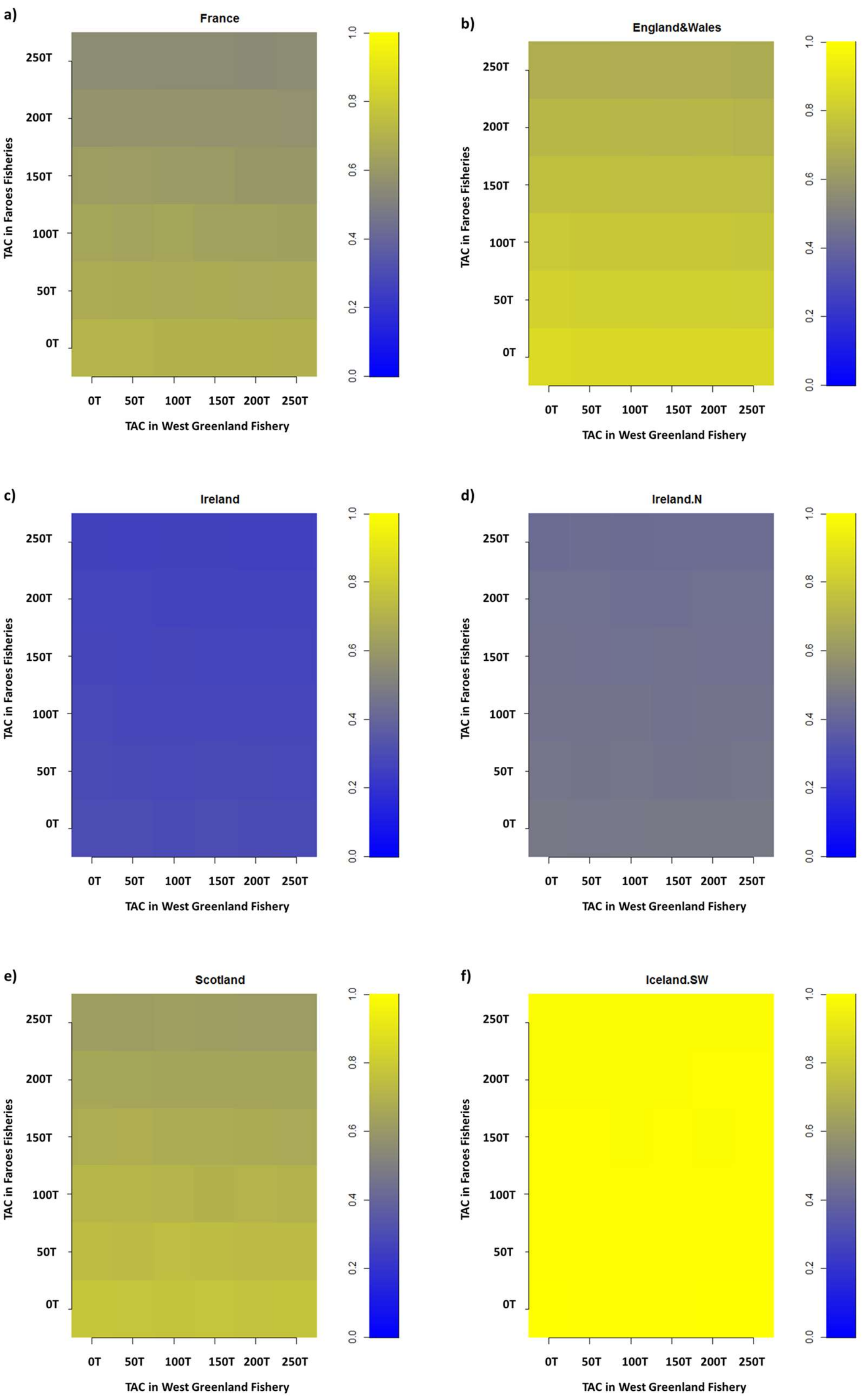

**Figure 19**. Probability to reach Conservation Limits simultaneously under different catches options at West Greenland and Faroes: 0, 50, 100, 150, 200 and 250 tons (5 years projections) for SU of the Southern European complex potentially impacted by both mixed stock fisheries.

# 6 Appendix 1. - Detailed process and observation equations of the Bayesian life cycle model

## 6.1 Population dynamics

### 6.1.1 Simplified life history

The age- and stage-structured life cycle model has a similar structure for each SU. It includes variation in the age of juveniles out-migrating from freshwater (i.e. smolts) and the sea-age of returning adults. Smolts migrate seaward after 1 to 6 years spent in freshwater (depending on SU). Two sea-age classes are considered in the model: Maiden salmon that return and reproduce after one year at sea, referred to as one-sea-winter (1SW) salmon or grilse, and maiden salmon that return after two winters spent at sea (2SW). This is a simplification of the variety of life history as some maiden fish may spent more than two winters at sea before returning to spawn, or some may be repeat spawners. However, those fish are rare and the 6 smolt-ages × 2 sea-ages combinations capture the essence of life history variations.

### 6.1.2 Eggs depositio*n*

The total number of eggs potentially spawned in year $t$ for SU $r$ is calculated from the number of 1SW ($N_{7_{t,r}}$) and 2SW ($N_{10_{t,r}}$) spawners escaping the homewater fisheries and the average number of eggs potentially spawned per 1SW and 2SW salmon, denoted $eggs_{1,r}$ and $eggs_{2,r}$ (fixed values; Table 2):

(A1.1) $$N_{1_{t,r}} = N_{7_{t,r}} \times eggs_{1,r} + N_{10_{t,r}} \times eggs_{2,r}$$

### 6.1.3 Egg-to-smolt transition

The egg-to-smolt transition consists of two steps: the survival from egg-to-smolt per cohort, and the distribution of the surviving smolts according to their age at downstream migration.

#### 6.1.3.1 Egg-to-smolt survival

Because no smolt production data is available at the scale of SU, it is difficult to separate the variability of the egg-to-smolt survival from that of the post-smolt survival, and parameters of the egg-to-smolt transitions have to be fixed. The egg-to-smolt survival is density independent, with average survival rate $\theta_1$ arbitrarily fixed to 0.007 (Hutchings & Jones, 1998; Massiot-Granier et al. 2014) for all years and all SU (Table 2). Environmental stochasticity is modelled by logNormal random noise with variance $\sigma_{\theta_1}{}^2$ fixed to an arbitrarily value corresponding to $CV_{\theta_1}$=0.4 ($\sigma_{\theta_1}{}^2 = \log(CV_{\theta_1}{}^2 + 1)$) which is a median values for the inter-annual variability found in the literature (Prevost et al., 2003; Pulkkinen et al., 2013). The total number of smolts produced in the cohort $c$ (corresponding to egg deposition of year $c$), denoted $N_{2_{c,r}}$ is then modelled as:

$$\text{(A1.2)} \qquad log\left(N_{2_{c,r}}\right) \sim Normal\left(log\left(\theta_1 * N_{1_{c,r}}\right) - \frac{1}{2}\sigma_{\theta_1}{}^2, \sigma_{\theta_1}{}^2\right)$$

This model configuration only allows for random stochasticity in the egg-to-smolt survival and does not account for any compensation neither (but see Olmos et al. 2019 for a sensitivity analysis to inclusion of density dependence). This implicitly assumes that any trends in the stock productivity over time are a response to changes in the marine phase, what may inflate the importance of trends in the post-smolt survival.

#### 6.1.3.2 Distribution according to smolt ages

The probabilities of a smolt in the cohort $c$ migrating at age $a = 1, \dots, 6$ at year $t = c + a + 1$, denoted $\theta_{2_{c,a,r}}$, are randomly drawn in tight informative Dirichlet priors with fixed averaged proportions $psm_{1:6,r}$ specific to each SU (Table 2):

$$\text{(A1.3)} \qquad \left(\theta_{2_{c,a=1,r}}, \dots, \theta_{2_{c,a=6,r}}\right) \sim Dirichlet\left(\eta_{sample} \times \left(psm_{1,r}, \dots, psm_{6,r}\right)\right)$$

The sample size of the Dirichlet distribution is arbitrarily fixed to $\eta_{sample} = 100$, and corresponds to the precision in the estimates of the proportions that would have been learned from multinomial samples of size $\eta_{sample}$.

Given $\theta_{2_{c,a,r}}$, the number of smolts from the cohort $c$ that migrate at age $a$ year $t = c + a + 1$ is modelled as:

(A1.4) $$log\left(N'_{2_{c,a,t=c+a+1,r}}\right) \sim Normal(log(\theta_{2_{c,a,r}} \times N_{2_{c,r}}) - \frac{1}{2}\sigma^2, \sigma^2)$$

with variance $\sigma^2$ fixed to an arbitrarily low value corresponding to CV=0.01.

Last, the number of smolts migrating in the spring of year $t$ is the sum of all smolts of different ages (and therefore of different cohorts) migrating in year $t$:

(A1.5) $$N_{3_{t,r}} = \sum_{a=1}^{a=6} N'_{2_{c=t-a-1,a,t,r}}$$

## 6.1.4 Marine phase

The marine phase is modelled as a sequence of three blocks of transitions: survival from smolts to the PFA stage, the maturation of fish at the PFA stage, and the fishing and natural mortality between PFA and returns.

### 6.1.4.1 Post-smolt survival and proportion of fish maturing as 1SW

Time series of post-smolt survivals ($\theta_{3_{t,r}}$) and the proportion of fish maturing as 1SW ($\theta_{4_{t,r}}$) are modelled as multivariate random walks in the logit scale. Random variations are drawn from multivariate Normal distributions with variance-covariance matrix $\Sigma_{\theta_3}$ and $\Sigma_{\theta_4}$ that define the covariations among the SU (Minto et al., 2014; Ripa and Lundberg, 2000):

(A1.6) $$\begin{cases} First\ year\ (for\ r = 1{:}N){:}\ logit\left(\theta_{3_{t=1,r}}\right) \sim Normal(0,1) \\ Then\ \left(logit\left(\theta_{3_{t+1,r}}\right)\right)_{r=1:N} \sim MVNormal\left(\left(logit\left(\theta_{3_{t,r}}\right)\right)_{r=1:N}, \Sigma_{\theta_3}\right) \end{cases}$$

(A1.7) $$\begin{cases} First\ year\ (for\ r = 1{:}N){:}\ logit\left(\theta_{4_{t=1,r}}\right) \sim Normal(0,1) \\ Then\ \left(logit\left(\theta_{4_{t+1,r}}\right)\right)_{r=1:N} \sim MVNormal\left(\left(logit\left(\theta_{4_{t,r}}\right)\right)_{r=1:N}, \Sigma_{\theta_4}\right) \end{cases}$$

Then, given the number of smolts migrating in year $t$ ($N_{3_{t,r}}$) and the post-smolt survival $(\theta_{3_{t,r}})$, the number of posts-smolts that survive to the PFA stage ($N_{4_{t+1,r}}$) in January of year $t+1$ is modelled as:

$$\text{(A1.8)} \qquad log\left(N_{4_{t+1,r}}\right) \sim Normal\left(log(\theta_{3_{t,r}} \times N_{3_{t,r}}) - \frac{1}{2}\sigma^2, \sigma^2\right)$$

Given the number of fish at the PFA stage ($N_{4_{t+1,r}}$) and the maturation rate ($\theta_{4_{t+1,r}}$), mature ($N_{5_{t+1,r}}$) and non mature fish ($N_{8_{t+1,r}}$) at the PFA stage are modelled as:

$$\text{(A1.9)} \qquad log\left(N_{5_{t+1,r}}\right) \sim Normal\left(log(\theta_{4_{t+1,r}} \times N_{4_{t+1,r}}) - \frac{1}{2}\sigma^2, \sigma^2\right)$$

$$\text{(A1.10)} \qquad log\left(N_{8_{t+1,r}}\right) \sim Normal\left(log((1-\theta_{4_{t+1,r}}) \times N_{4_{t+1,r}}) - \frac{1}{2}\sigma^2, \sigma^2\right)$$

### 6.1.4.2 Sequential marine fisheries and natural mortality

After the PFA stage, both maturing and non-maturing fish are subject to natural mortality and sequential fisheries mortalities operating on mixed stocks (Tables 4 & 5). The following modelling structure applies for each of those transitions. For any marine fishery $f$, operating in year $t$ on a number of fish $N_{f_{t,r}}$ originated from the stock unit $r$ with an exploitation rate $h_{f_{t,r}}$, the catches $C_{f_{t,r}}$ (unknown states) and the number of fish that escape the fishery $N_{f.esc\ t,r}$ are modelled as:

$$\text{(A1.11)} \qquad C_{f_{t,r}} = h_{f_{t,r}} \times N_{f_{t,r}}$$

$$\text{(A1.12)} \qquad N_{f.esc\ t,r} = (1 - h_{f_{t,r}}) \times N_{f_{t,r}}$$

Exploitation rates $h_{f_{t,r}}$ are modelled as variable over time but their variability across SU is modelled differently depending on the data available to allocate catches to each SU and on expert knowledge about migration routes. Exploitation rates of the West Greenland fishery (WG; operating on a mixture of SU from North America and Europe) and of the Faroes fishery (FA; operating on SU from Europe only) were all supposed to vary across years and SU (Tables 4 & 5). For the fisheries specific to the SU from NA (Table 4), catches were allocated to each SU by considering a single $h$ homogeneous for all SU. There are two exceptions to this general rule (Prévost et al., 2009). The first is for the Labrabor/Newfoundland (LAB/NFDL) fishery on 1SWm and 2SW fish for which a separate $h$ is estimated for Labrador and one single $h$ is considered for the five other SU. A second exception is for the Saint-Pierre et Miquelon (SPM) fishery on 1SWm and 2SW for which $h$ of fish originating from Labrador was fixed to zero for all years.

All fisheries at sea are separated by periods of time where only natural mortality occurs (ICES, 2015a; Potter, 2016; Prévost et al., 2009). Fish that escape the fishery $f$ at year $t$ hence suffer natural mortality rate $\theta_{5_{t,f}} = \mathrm{e}^{-\mathrm{M}\times\Delta_{\mathrm{t,f}}}$ where the monthly mortality rate $M$ is fixed, constant across years and SU's ($M = 0.03 \cdot month^{-1}$ ; Table 2) and the duration $\Delta_{\mathrm{t,f}}$ (in months) are assumed known and constant across years but with some variations among SU to account for variability in migration routes (Tables 4 & 5):

$$\text{(A1.13)} \qquad N_{f+1\ t,r} = (1 - \theta_{5_{t,f}}) \times N_{f.esc\ t,r}$$

### 6.1.4.3 From returns to spawners (homewater catches)

Fish that escape all marine mortality and return as 1SW fish ($\mathrm{N}_{6_{\mathrm{t,r}}}$) or 2SW fish ($\mathrm{N}_{9_{\mathrm{t,r}}}$), are subject to homewater fisheries that operate locally on each SU. Homewater fisheries are modelled with exploitation rates $\mathrm{h}_{\mathrm{HWf}_{\mathrm{t,r}}}$ that are assumed to vary with years and SU and for the two sea-age classes separately (Tables 5 & 5). Homewater fishery harvest rates are estimated. After homewater fishery, a proportion of fish may potentially delay spawning to the next year. The proportion of delayed spawners are supposed known but varies with SU, years and sea-age classes and are denoted $p_{delSp_{t,r}}$. Fish that delay spawning to the next year may then be subject to a specific fishery with (estimated) harvest rates $h_{delSp_{t,r}}$. In practice, the proportion of delayed spawners is non-zero only for Russian stock units. But the transitions are modelled uniformly for all stock units with zero proportion of delayed spawners in the data for almost all SU. Last, the number of 2SW spawners in the US stock unit is also supplemented by stocking. The transition is also modelled uniformly for all SU but the number of fish stocked $n_{\mathrm{Stock.2SW}_{t,r}}$ is null for all SU except USA. Finally, the number of fish that escape the homewater fishery and potentially spawn as 1SW ($\mathrm{N}_{7_{\mathrm{t,r}}}$) and 2SW ($\mathrm{N}_{10_{\mathrm{t,r}}}$) are modelled as:

(A1.14)

$$N_{7\,t,r} = \left(1 - h_{HWf.1SW_{t,r}}\right) \times \left(1 - p_{delSp.1SW_{t,r}}\right) \times N_{6_{t,r}} + \left(1 - h_{HWf.1SW_{t-1,r}}\right) \times p_{delSp.1S\ \ t-1,r} \times \left(1 - h_{delSp.1SW_{t,r}}\right) \times N_{6_{t-1,r}}$$

(A1.15)

$$N_{10\,t,r} = \left(1 - h_{HWf.2SW_{t,r}}\right) \times \left(1 - p_{delSp.2SW_{t,r}}\right) \times N_{9_{t,r}} + \left(1 - h_{HWf.2SW_{t-1,r}}\right) \times p_{delSp.2SW_{t-1,r}} \times \left(1 - h_{delSp.2SW_{t,r}}\right) \times N_{9_{t-1,r}} + n_{\mathrm{Stock.2SW}_{t,r}}$$

## 6.2 Observation equations

The model incorporates observation errors for the time series of returns and catches. A sequential approach (Michielsens et al., 2008; Staton et al., 2017) is used that consists of two steps:

- In a first step, observation models are processed separately to reconstruct probability distributions that synthetize observation uncertainty around catches and returns for each year and each of the 24 SU. Probability distributions for returns and catches are derived from a variety of raw data and observation models, specific to each SU and each year and originally developed by ICES to provide input for PFA models for NA (Rago et al., 1993) and SE (Potter et al., 2004b) stock units.
- In a second step, those distributions are used to approximate likelihoods in the population dynamics state-space model.

### 6.2.1 Returns

Returns are estimated for each year, each SU and for the two sea-age classes separately. Raw data used to estimate return essentially consist in homewater catches available at the scale of rivers or regional fishery jurisdictions, scaled by harvest and declaration rates and then aggregated at the scale of larger stock units. Uncertainties then essentially arise from a numerical (Monte Carlo) integration of uncertainties about harvest and declaration rates. Other fishery independent information like counting fences or mark and recapture data can also be used. Detailed description of the raw data and models used in each SU is provided in the WGNAS Stock Annex for Atlantic salmon (Crozier et al., 2003; ICES, 2002, 2015b; Potter et al., 2004b; Rago et al., 1993).

#### 6.2.1.1 The case of Northern NEAC SU

ICES provides a shorter time series of data for Northern NEAC SU because some data are missing for Norway for the first time of the time series before 1982. The Norwegian data for the period 1971-1982 were completed using the following hypotheses (*Com pers*. Geir Bolstad and Peder Fiske, NINA):

- Homewater catches - Catch data for Norway (homewater catches, 1SW and 2SW separately) for the period 1971-1982 were extracted from the ICES WGNAS report of year 2002 (table 3.3.3.1f. Allocations of catches among the four regions of Norway was done using averages proportions calculated from the five previous years for which data are available 1983-1987.
- Returns – The probability distribution of returns (1SW and 2SW, separately) was estimated by dividing the catches by guest estimates of exploitation rates and unreported catches for the period 1982-1971. Harvest rates and unreported catches were extrapolated backwards in time from year 1983. Uncertainty about those rates was bumped by 20% to account for the additional uncertainty due to extrapolation.
- Note that all MSW were considered as 2SW as for all other European SU .

The resulting probability distributions of returns are shown in Fig. 4. Numerical integration of uncertainty support the hypothesis that the returns are logNormaly distributed, allowing to approximate the likelihood for the returns as follows. For any year $t$ and SU $r$, the expected mean of the distribution derived from the observations models for 1SW (respectively, 2SW) returns in log scale, denoted $\mathbb{E}_{log\left(R_{1SW_{t,r}}\right)}$ (resp. $\mathbb{E}_{log\left(R_{2SW_{t,r}}\right)}$), is considered as a observed realization of a Normal distribution of non-observed returns (in log-scale) $\mathrm{N}_{6_{t,r}}$ (resp. $\mathrm{N}_{9_{t,r}}$), with known variance $\sigma^2_{1SW_{t,r}}$ (resp. $\sigma^2_{2SW_{t,r}}$ ) set to the value derived from the observation errors models. These observation errors are considered independent across years, SU and sea-age classes.

(A1.16) $\mathbb{E}_{log\left(R_{1SW_{t,r}}\right)} \sim Normal\left(log\left(N_{6_{t,r}}\right), \sigma^2_{1SW_{t,r}}\right)$

(A1.17) $\mathbb{E}_{log\left(R_{2SW_{t,r}}\right)} \sim Normal\left(log\left(N_{9_{t,r}}\right), \sigma^2_{2SW_{t,r}}\right)$

### 6.2.2 Homewater catches

The homewater fisheries take adult fish that are mainly returning to the natal rivers to spawn. Point estimates of total catches reported by ICES (ICES 2015b) pool all homewater fisheries capturing returning fish in coastal areas, estuaries and freshwater, for each SU, each year and each sea-age class separately (Fig. 5). They are here denoted $C_{\mathrm{HW}.1SW_{t,r}}$ and $C_{\mathrm{HW}.2SW_{t,r}}$ for 1SW and 2SW fish, respectively. The likelihood term for homewater catches is built from logNormal observation errors with known observation error. Available knowledge support that homewater catches are known with only few errors. Relative error is then arbitrarily fixed to CV=0.05 for both sea-ages, for all years and all SU. Observation errors are considered independent across years, SU and sea-age classes. The likelihood terms associated with homewater catches are:

(A1.18)

$$\log\left(C_{\mathrm{HW}.1S\ \ t,r}\right) \sim Normal\left(\log\left(h_{HWf.1SW_{t,r}} \times \left(1 - p_{delSp.1S\ \ t,r}\right) \times N_{6_{t,r}}\right), \sigma^2_{\mathrm{HW.1SW}}\right)$$

(A1.19)

$$\log\left(C_{\mathrm{HW}.2SW_{t,r}}\right) \sim Normal\left(\log\left(h_{HWf.2SW_{t,r}} \times \left(1 - p_{delSp.2SW_{t,r}}\right) \times N_{9_{t,r}}\right), \sigma^2_{\mathrm{HW.2SW}}\right)$$

with $\sigma^2_{\mathrm{HW.1SW}} = \sigma^2_{\mathrm{HW.2SW}}$ the variance associated with CV=0.05.

Observation model for the delayed catches are modelled using the same hypothesis and the same CV of observation errors.

### 6.2.3 Catches at sea for sequential distant marine fisheries operating on mixed stocks

For any marine fishery $f$ operating on a mixture of SU, likelihood equations consist in logNormal observation errors on the total catches summed over all SU (still based on the same likelihood approximation method), eventually supplemented by Dirichlet likelihood terms to adjust the proportion of catches allocated to each SU when proportion data are available (Table 6 and Fig. 6 & 8). Observation errors on the total catches and on the proportions are considered independent across fisheries, years and SU.

Observation models based on ICES (2017b) data are built independently from the state-space model to estimate logNormal probability distributions of total catches at sea for each fishery $f$ and each year $t$, with expected mean and variance (in log-scale) denoted $\mathbb{E}_{\log(C_{f_t})}$ and ${\sigma^2}_{\mathrm{f_t}}$, respectively. Variances ${\sigma^2}_{f_t}$ are derived by integrating uncertainty in the catch declaration rates, the proportions of fish of wild origin in the catches, and sampled biological characteristics of the catches including average weight of a fish used to convert catches in weights to number of fish, and scale samples used to separate the two sea-age classes in the catches. An exception is for the WG fishery for which observation errors are considered to be low (ICES 2005b) and fixed to $CV = 0.1$.

By denoting $C_{f_t} = \sum C_{f_{t,r}}$ is the total catches from the state process summed over all SU, the likelihood term for the total catch is modelled as:

$$(\mathrm{A1.20}) \qquad \mathbb{E}_{log(C_{f_t})} \sim Normal\big(log(C_{f_t}), {\sigma^2}_{f_t}\big)$$

Proportion of catches allocated to each SU are available for the West Greenland fishery (European and North American continental stock groupings) and for the Faroes fishery (1SWm and 1SWnm, and 2SW, for the European continental stock groupings only).

Proportions used to allocate West Greenland catches to each of the 24 SU in North America and Europe (Fig. 6) are derived from a compilation of individual assignment data from scale reading and genetic analyses. Proportions of the total catches at WG are first attributed to European and North American based on scales (1971-1999) and genetics samples (2000-2014) (ICES 2017a; ICES 2017b). Then, proportions attributed to each SU within the European stock group are fixed through time as compiled from ICES (2017b). Within the North American continental stock group, proportions are based on Bradbury et al. (2016a,2016b) that provide estimates of the proportion of fish originated from North American SU for 13 years based on genetic samples. The average value of the 13 years are used for the years without available data.

Proportions used to allocate Faroes catches to European SU are derived from a compilation of assignment data from scale reading (to separate fish from Southern and Northern Europe origin) and genetics data to allocate to each SU (ICES, 2015a). Data are not informative enough to account for annual variability and those proportions are considered constant over the time series (Table 6, Fig. 8).

When available, observed proportion of each SU in the total catches, denoted $p^{obs}_{f_{t,r}}$ enters into a Dirichlet likelihood modelled as:

$$\text{(A1.21)} \qquad \left(p^{obs}_{f_{t,r=1}}, \dots, p^{obs}_{f_{t,r=K}}\right) \sim Dirichlet\left(\eta_{sample} \times \left(p_{f_{t,r=1}}, \dots, p_{f_{t,r=K}}\right)\right)$$

where $p_{f_{t,r}} = \frac{C_{f_{t,r}}}{C_{f_t}}$ is the proportion of fish from SU $r$ in the total catches calculated from the state process. When no proportions data are available, only the logNormal likelihood on total catches is used. The hypothesis of a homogeneous exploitation rate among SU replaces the Dirichlet likelihood. As a direct consequence, the proportions of any SU in the catches are set in pro-rata to the abundance among the SU just before the fishery.